\documentstyle[emulateapj]{article}

\lefthead{Nagao, Murayama, and Taniguchi}
\righthead{The NLR of Seyferts: NLS1s vs. BLS1s}

\begin{document}

\submitted{Accepted for publication in the Astrophysical Journal}
\title{The Narrow-Line Region of Seyfert Galaxies: \\
       Narrow-Line Seyfert 1s versus Broad-Line Seyfert 1s}

\author{Tohru NAGAO, Takashi MURAYAMA, and Yoshiaki TANIGUCHI}
\affil{Astronomical Institute, Graduate School of Science, Tohoku University,
       Aramaki, Aoba, Sendai 980-8578, Japan\\
       tohru@astr.tohoku.ac.jp, murayama@astr.tohoku.ac.jp, 
       tani@astr.tohoku.ac.jp}

\begin{abstract}

It is known that
the spectral energy distribution (SED) of the nuclear radiation of
narrow-line Seyfert 1 galaxies (NLS1s) has different shapes with respect to
that of ordinary broad-line Seyfert 1 galaxies (BLS1s),
particularly in wavelengths
of X-ray. This may cause some differences in the ionization degree and the 
temperature of gas in narrow-line regions (NLRs)
between NLS1s and BLS1s.
This paper aims to examine whether or not there are such differences in the
physical conditions of NLR gas between them.
For this purpose, we have compiled the emission-line ratios
of 36 NLS1s and 83 BLS1s from the literature.
Comparing these two samples, we have found that the line ratios of
[O {\sc i}]$\lambda$6300/[O {\sc iii}]$\lambda$5007 and 
[O {\sc iii}]$\lambda$4363/[O {\sc iii}]$\lambda$5007, which represent
the ionization degree and the gas temperature respectively, are
statistically indistinguishable between NLS1s and BLS1s.
Based on new photoionization model calculations, we show
that these results are not inconsistent with the
difference of the SED between them.
The influence of the difference of SEDs on the highly ionized emission lines
is also briefly discussed.

\end{abstract}


\section{INTRODUCTION}

Seyfert nuclei are typical active galactic nuclei (AGNs) 
in the nearby universe.
They have been broadly classified into two types based on 
presence or absence of broad emission lines in their optical spectra
(Khachikian \& Weedman 1974); Seyferts with broad lines are type 1 
(hereafter S1) while those without broad lines are type 2 (S2). These two 
types of Seyfert nuclei are now unified by introducing the viewing angle 
dependence toward the central engine surrounded by the geometrically and 
optically thick dusty torus (Antonucci \& Miller 1985; see for a review 
Antonucci 1993).

In addition to these typical types, narrow-line Seyfert 1 galaxies
(NLS1s) have also been recognized as a distinct type of Seyfert
nuclei. The optical emission-line properties of NLS1s are summarized as 
follows (e.g., Osterbrock \& Pogge 1985). (1) The Balmer lines are only 
slightly broader than the forbidden lines such as [O {\sc iii}]$\lambda$5007
(typically less than 2000 km s$^{-1}$). This property makes NLS1s a distinct 
type of ordinary broad-line S1s (BLS1s). 
(2) The [O {\sc iii}]$\lambda$5007/H$\beta$ intensity
ratio is smaller than 3. This criterion was introduced to discriminate S1s 
from S2s by Shuder \& Osterbrock (1981). And, (3) They present strong 
Fe {\sc ii} emission lines 
which are often seen in S1s but generally not in S2s.
Moreover, the X-ray spectra of NLS1s are very steep (Puchnarewicz et al. 1992;
Boller, Brandt, \& Fink 1996; Wang, Brinkmann, \& Bergeron 1996;
Vaughan et al. 1999; Leighly 1999b) and
highly variable (Boller et al. 1996; Turner et al. 1999; Leighly 1999a).
Because of these complex properties, it is still not understood what NLS1s are
in the context of the current AGN unified model.

In order to understand what NLS1s are, it is important to investigate the 
narrow-line regions (NLRs) of NLS1s because of the following two reasons.
First, the intrinsic spectral energy distribution (SED) of 
the nuclear radiation of NLS1s is rather different from that of BLS1s;
i.e., the soft and hard
X-ray spectra of NLS1s are steeper than those of BLS1s (Boller et al. 1996;
Brandt, Mathur, \& Elvis 1997;
Vaughan et al. 1999; Leighly 1999b). 
It is often considered that the NLRs are photoionized by the nonthermal 
continuum radiation from central engines (Yee 1980; Shuder 1981; Cohen 1983;
Cruz-Gonz\'{a}lez et al. 1991; Osterbrock 1993; Evans et al. 1999)
though shock ionization may play an important role of ionization of the NLR
(e.g., Contini \& Aldrovandi 1983; Viegas-Aldrovandi \& Contini 1989;
Dopita \& Sutherland 1995). If the dominant mechanism of ionization is the
photoionization, the degree or the structure of ionization of NLRs in NLS1s
may be different from that of BLS1s.
Such difference can be probed by forbidden emission-line ratios.
Second, it is known that there are some differences between the NLR properties
of S1s and S2s, for example, the gas temperature in the [O {\sc iii}] zone
(e.g., Heckman \& Balick 1979; Shuder \& Osterbrock 1981).
Although the reason of such differences has not yet been understood fully, 
it is meaningful to investigate how the NLRs in NLS1s share the properties
with those in S1s or those in S2s.

Analyzing optical spectra of 7 NLS1s and 16 BLS1s,
recently, Rodr\'{\i}guez-Ardila, Pastoriza, \& Donzelli (2000b) and 
Rodr\'{\i}guez-Ardila et al. (2000a) reported that the NLRs of NLS1s are
less excited than those of the BLS1s.
They suggested that this is due to the difference in the shape of the 
SEDs of nuclear radiation between NLS1s and BLS1s.
In their analysis they used the intensities of the forbidden lines 
normalized by the narrow components of Balmer lines.
However, it is not clear whether or not
the ``narrow components'' of the Balmer lines of NLS1s are radiated from 
only the NLRs. 
For example, line widths of the Balmer lines radiated from broad-line 
regions (BLRs) may be narrow like NLR emission if we see NLS1s from a
nearly pole-on viewing angle (Taniguchi, Murayama, \& Nagao 1999 and
reference therein).
Therefore it seems better to use some combinations among forbidden
emission lines.

In this paper, we present the comparisons of some emission-line
flux ratios among NLS1s and BLS1s (see Nagao, Taniguchi,
\& Murayama 2000c for highly ionized emission lines) using the data compiled 
from the literature. 

\section{DATA COMPILATION}

 \subsection{Data}

In order to investigate the properties of the NLRs in NLS1s and in BLS1s,
we compiled the following emission lines from the literature; 
[O {\sc i}]$\lambda$6300, [O {\sc ii}]$\lambda\lambda$3726,3729,
[O {\sc iii}]$\lambda$5007, [O {\sc iii}]$\lambda$4363, 
[N {\sc ii}]$\lambda$6583, and [S {\sc ii}]$\lambda\lambda$6717,6731.
These lines are respectively referred as [O {\sc i}], [O {\sc ii}],
[O {\sc iii}]$\lambda$5007, [O {\sc iii}]$\lambda$4363, [N {\sc ii}],
and [S {\sc ii}] below. As mentioned in Section 1, we do not use
the flux of the Balmer lines in order to avoid any ambiguity.
The number of compiled objects is 119; 36 NLS1s and 83 BLS1s.
The so-called Seyfert 1.2 galaxies
(see Osterbrock 1977 and Winkler 1992) are also included in the BLS1 sample.

All the Seyfert galaxies are listed up in Table 1 together with their 
redshifts and 60$\mu$m luminosities\footnote{
   In this paper, we adopt a Hubble constant {\it H}$_{\rm 0}$ = 
   50 km s$^{-1}$ Mpc$^{-1}$ and a deceleration parameter 
   {\it q}$_{\rm 0}$ = 0.
}.
The 60 $\mu$m luminosities are taken from the {\it IRAS} Faint Source Catalogue
(Moshir et al. 1992). 
The emission-line flux ratios for each object are given in Table 2.
Each ratio is the averaged value among the references given in Table 1.
Since it is often difficult to measure the narrow Balmer component for S1s
accurately, there might be the systematic error if we make reddening
corrections using the Balmer decrement method (e.g., Osterbrock 1989) 
for both types of Seyferts. Therefore, we do not make the
reddening correction.
The effect of dust extinction on our results is discussed in Section 3.3.

Some galaxies do not have all the emission-line ratios.
The lack of the data in Table 2 is attributed to the following
five reasons; (1) the observation did not cover the wavelengths where 
the emission lines exist, (2) the emission-line was not detected and 
the upper limit is not given in the reference, (3) the only upper limit is 
given in the reference, (4) the flux of the emission lines were not 
given in the reference because the author(s) of the reference were not 
interested in those emission lines, and (5) the de-blending of 
the [O {\sc iii}]$\lambda$4363 emission from the H$\gamma$ emission 
and the [N {\sc ii}]$\lambda$6583 emission from the H$\alpha$ emission 
were not performed in the reference. 
Since the number of the upper-limit data is quite
small and those values are too large to be used for any scientific 
discussion, we do not use these upper-limit data in later analyses.

 \subsection{Selection Bias}

\begin{figure*}
\epsscale{0.6}
\plotone{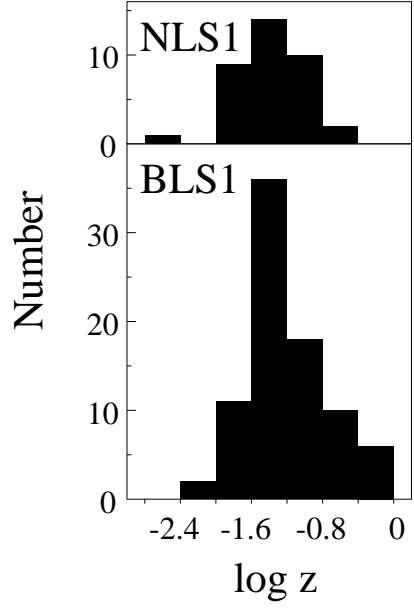}
\caption{
The frequency distributions of the redshift for the NLS1s and the BLS1s.
\label{fig1}}
\end{figure*}

\begin{figure*}
\epsscale{0.6}
\plotone{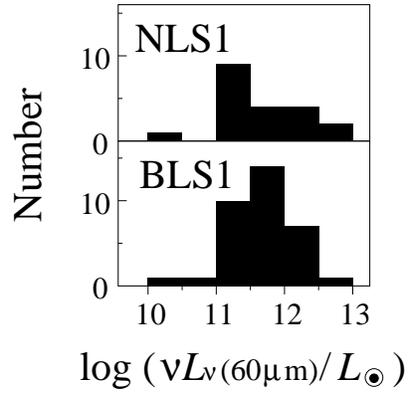}
\caption{
The frequency distributions of the 60$\mu$m luminosity 
for the NLS1s and the BLS1s.
The luminosities are normalized by the solar luminosity.
\label{fig2}}
\end{figure*}

\begin{figure*}
\epsscale{1.2}
\plotone{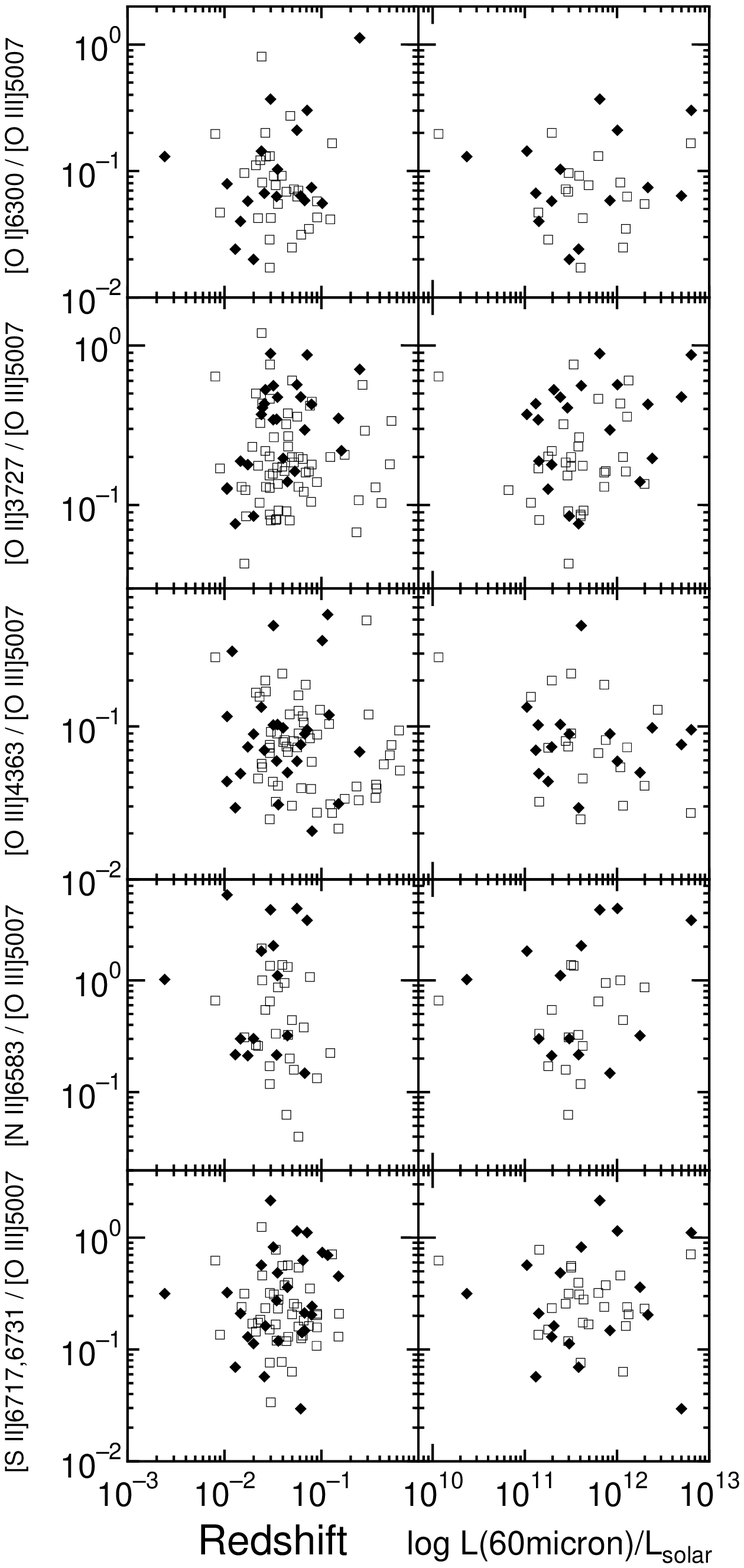}
\caption{
Line ratios described in text are plotted as functions of redshift and
the 60$\mu$m luminosity.
The compiled data of the NLS1s and the BLS1s are
shown by filled diamonds and open squares, respectively.
\label{fig3}}
\end{figure*}

Because we do not impose any selection criteria upon our samples,
it is necessary to check whether or not the two samples are 
appropriate for our comparative study.
If there are some systematic differences in the redshift distribution 
and in the intrinsic AGN power distribution between the two samples,
there would be possible biases.

First we investigate the redshift distribution.
We show the histograms of the redshift in Figure 1.
The average redshifts and 1$\sigma$ deviations are
0.0568 $\pm$ 0.0502 for the NLS1s and 0.1141 $\pm$ 0.1455 for the BLS1s.
It is noted that the average redshift of the BLS1s
is a little higher than that of the NLS1s.
In order to investigate whether or not the frequency distributions of
the redshift are statistically different between two samples, 
we apply the Kolmogorov-Smirnov (KS) statistical test (Press et al. 1988).
The null hypothesis is that the redshift distributions of the NLS1s and 
the BLS1s come from the same underlying population.
The resultant KS probability is 4.650 $\times$ 10 $^{-1}$,
which means that the two distributions are statistically indistinguishable.
Hence we conclude that there is no redshift bias.

Second, we investigate whether or not the intrinsic 
AGN power is systematically different between the two samples
using the {\it IRAS} 60$\mu$m luminosity, 
which is regarded as a rather isotropic emission (Pier \& Krolik 1992; 
Efstathiou \& Rowan-Robinson 1995; Fadda et al. 1998).
The 60$\mu$m luminosity is thought 
to scale the nuclear continuum radiation which is absorbed and re-radiated by 
the dusty torus (see Storchi-Bergmann, Mulchaey, \& Wilson 1992). 
The histograms of the 60$\mu$m luminosity are shown in Figure 2.
The average 60$\mu$m luminosities and 1$\sigma$ deviations
in logarithm (in units of solar luminosity) are
11.646 $\pm$ 0.646 for the NLS1s and 11.627 $\pm$ 0.487 for the BLS1s.
We apply the KS test where the null hypothesis is that the distribution of
the 60$\mu$m luminosity of the two samples come from the same underlying 
population. The resultant KS probability is 3.884 $\times 10 ^{-1}$, 
which means that there is no
systematic difference of the 60$\mu$m luminosity between two samples.
Although the 60$\mu$m luminosity might be contaminated with the influence of
circumnuclear star formation and have the weak unisotropic tendency,
this test supports the validity of the statistical comparisons in our study.

In Figure 3 we also show
that the line ratios which we compiled do not correlate with
the redshift and the 60$\mu$m luminosity.

\section{COMPARISON OF LINE RATIOS}

 \subsection{The Ionization Degree of the NLRs}

\begin{figure*}
\epsscale{1.0}
\plotone{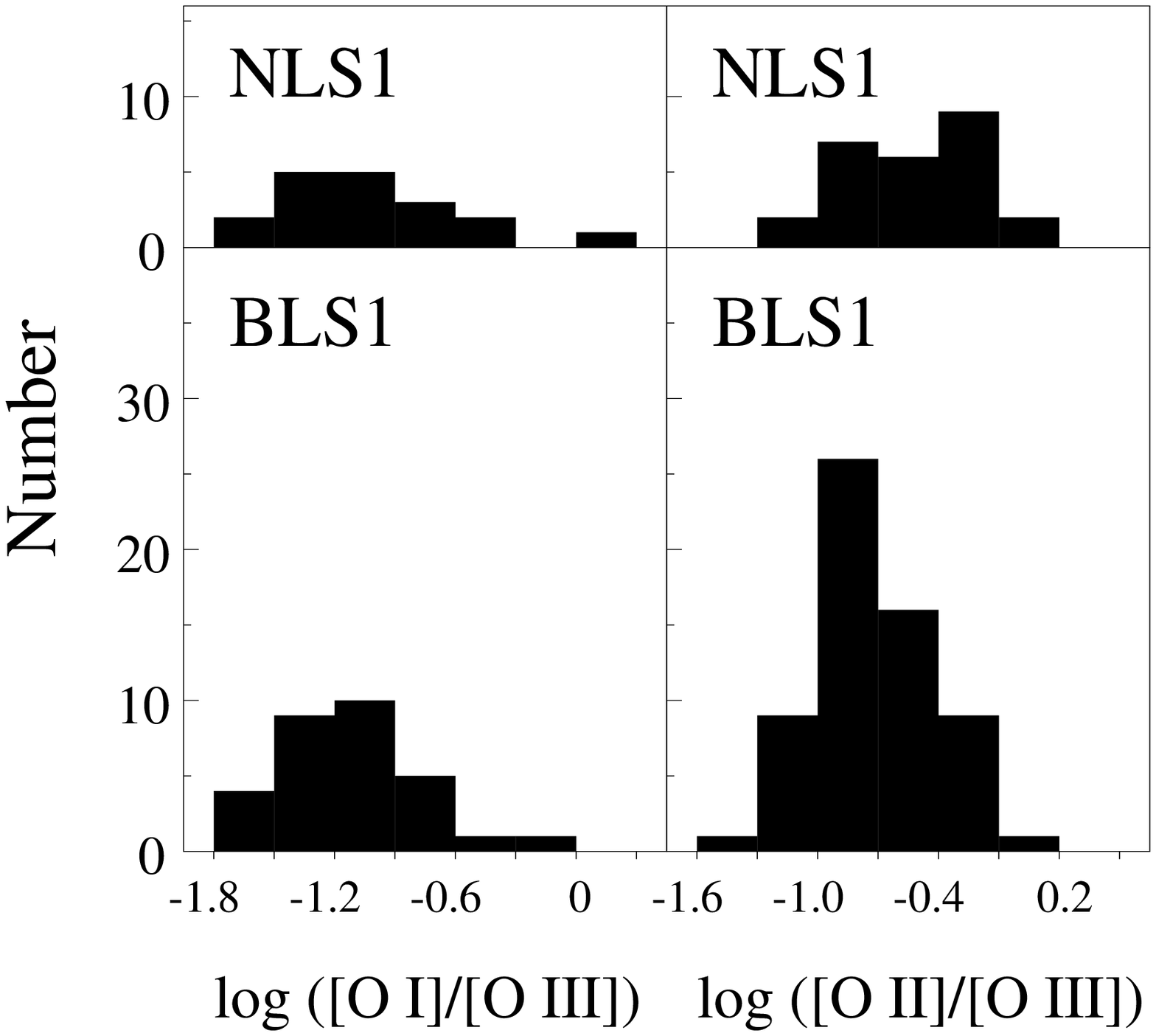}
\caption{
The frequency distributions of the line ratios of [O {\sc i}] and 
[O {\sc ii}] to [O {\sc iii}]$\lambda$5007
for the NLS1s and the BLS1s.
\label{fig4}}
\end{figure*}

To investigate whether or not the ionization degree of the
NLRs is different between NLS1s and BLS1s,
we compare some emission-line ratios between the two samples.
Although emission-line ratios of AGNs have been traditionally discussed in the
form normalized by a narrow component of Balmer lines, for example
[O III]$\lambda$5007/H$\beta$, we do not use such line ratios
because of the difficulty in deblending of the narrow
component of Balmer lines from the broad component.

In this study, we investigate the ionization degree of NLRs using [O {\sc i}], 
[O {\sc ii}], and [O {\sc iii}]$\lambda$5007. 
The ionization potentials of the lower stage of ionization
(to produce the relevant ions) are 0.0 eV, 13.6 eV, and 35.1 eV, respectively.
This set is free from the chemical abundance effect.
Here we use [O {\sc i}]/[O {\sc iii}]$\lambda$5007 and 
[O {\sc ii}]/[O {\sc iii}]$\lambda$5007.
It is noted that the critical density of [O {\sc iii}]$\lambda$5007 is 
similar to that of [O {\sc i}] (7.0 $\times$ 10$^5$ cm$^{-3}$ and 1.8 
$\times$ 10$^6$ cm$^{-3}$, respectively) and far larger than that of
[O {\sc ii}] (4.5 $\times$ 10$^3$ cm$^{-3}$).
This may mean that [O {\sc i}] and [O {\sc iii}]$\lambda$5007 are 
radiated from the similar region in NLRs while the [O {\sc ii}] emission
comes from relatively lower-density clouds.
For example, it may be such the case that matter-bounded parts of a 
relatively high-density cloud radiate the [O {\sc iii}]$\lambda$5007
emission while ionization-bounded parts of the same cloud radiate the
[O {\sc i}] emission (see Figure 4b of Binette, Wilson, \& 
Storchi-Bergmann 1996).
Therefore [O {\sc i}]/[O {\sc iii}]$\lambda$5007 seems better to
investigate the ionization degree of gas clouds in NLRs than 
[O {\sc ii}]/[O {\sc iii}]$\lambda$5007.

We show the histograms of these emission-line ratios for the NLS1s and
the BLS1s in Figure 4.
Though the NLS1s appear to have larger [O {\sc ii}]/[O {\sc iii}]$\lambda$5007
than the BLS1s, it seems that there is no systematic difference in
[O {\sc i}]/[O {\sc iii}]$\lambda$5007 between the two samples.
In order to investigate whether or not these distributions of the
emission-line ratios are statistically different between the two samples,
we apply the KS test.
The null hypothesis is that the distributions of the relevant ratio of
the samples come from the same underlying population.
The KS probabilities are 7.089 $\times 10^{-1}$ for
[O {\sc i}]/[O {\sc iii}]$\lambda$5007 and 6.318 $\times 10^{-3}$ for
[O {\sc ii}]/[O {\sc iii}]$\lambda$5007,
which lead to the following results.
1) There is no statistical difference in [O {\sc i}]/[O {\sc iii}]$\lambda$5007
between two samples. 2) It is, however, not clear whether or not 
there is statistical 
difference in [O {\sc ii}]/[O {\sc iii}]$\lambda$5007 between them.
These results seem to suggest that there is little difference of the 
ionization degree of the NLRs between the two types of Seyferts.
If this is true,
it is contradictory to the result of Rodr\'{\i}guez-Ardila et al. (2000b),
who reported that NLS1s are less excited objects than BLS1s.
To make this issue clear, we have carried out the model calculations,
which is presented in Section 4.

 \subsection{The [O {\sc iii}] Line Ratio}

\begin{figure*}
\epsscale{0.6}
\plotone{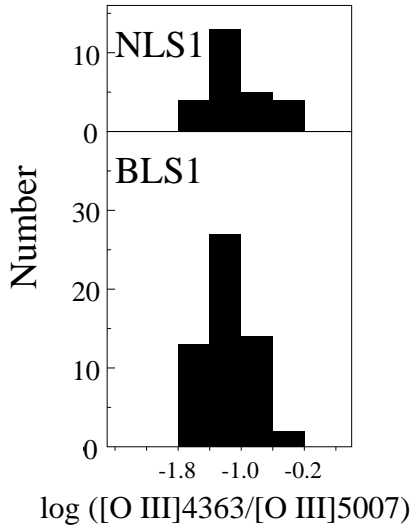}
\caption{
The frequency distributions of 
[O {\sc iii}]$\lambda$4363/[O {\sc iii}]$\lambda$5007
for the NLS1s and the BLS1s.
\label{fig5}}
\end{figure*}

We investigate the 
[O {\sc iii}]$\lambda$4363/[O {\sc iii}]$\lambda$5007 ratio, which is
sensitive to the gas temperature (e.g., Osterbrock 1989). 
In Figure 5, we show the histograms of
[O {\sc iii}]$\lambda$4363/[O {\sc iii}]$\lambda$5007 for the NLS1s and
the BLS1s.
In order to investigate whether or not 
these distributions of both samples are statistically different,
we also apply the KS test where the null hypothesis is that 
the distributions of the emission-line ratio between two samples
come from the same underlying population.
The KS probability is 7.877 $\times 10 ^{-1}$, 
which means that there is no statistical difference in this line ratio 
between the NLS1s and the BLS1s.

 \subsection{The Effects of the Dust Extinction}

As mentioned in Section 2.1, no reddening correction has been made for 
all the collected emission-line ratios analyzed here.
However, it is known that dust grains are present in the NLR of Seyferts
(e.g., Dahari \& De Robertis 1988a, 1988b; Netzer \& Laor 1993).
Hence we check how the extinction affects the emission-line
ratios discussed in previous sections.
Because the difference of the average amounts of the extinction between
S1s and S2s is about 1 magnitude (Dahari \& De Robertis 1988a; see
also De Zotti \& Gaskell 1985), we investigate the extinction effect
in the case of $A_V$ = 1.0 mag using the Cardelli's extinction curve
(Cardelli, Clayton, \& Mathis 1989).
Correction factors for the observed values of 
[O {\sc i}]/[O {\sc iii}]$\lambda$5007, 
[O {\sc ii}]/[O {\sc iii}]$\lambda$5007, and
[O {\sc iii}]$\lambda$4363/[O {\sc iii}]$\lambda$5007 for the extinction
($A_V$ = 1.0 mag) are 0.786, 1.471, and 1.222, respectively.
These values correspond to about a half bin in Figures 3, 4, and 5.
This suggests that the extinction might affect the results 
in Sections 3.1 and 3.2 
$if$ there is a systematic difference in the amounts of the extinction 
much more than 1 magnitude between NLS1s and BLS1s.
However, the sample of Rodr\'{\i}guez-Ardila et al. (2000b) showed
little difference in the amounts of the extinction between
NLS1s and BLS1s: $A_V$ = 0.457 $\pm$ 0.137 for 7 NLS1s and 0.663 $\pm$ 0.345
for 16 BLS1s. 
Though the number of objects is small, this suggests that the difference
in the amounts of the extinction is so small that the extinction does not
affect the results presented in previous sections.

\section{MODEL CALCULATIONS}

Now we must consider the following problem. 
It has been known that the shape of SEDs of nuclear radiation is different
between NLS1s and BLS1s particularly in X-ray band.
Since UV to X-ray photons are closely connected with the photoionization 
process, such difference in the SED may cause some distinctions
in physical properties of the ionized gas in NLRs, such as
the ionization degree and the temperature.
On the other hand,
our comparative study described in Section 3 suggests that there is
little difference in the ionization degree and in the temperature of the 
gas in NLRs between NLS1s and BLS1s.
Is this result plausible in terms of photoionization models?
In order to investigate this issue, we carry out photoionization 
model calculations
and compare the model results with the compiled 
emission-line ratios.

 \subsection{The SEDs of NLS1s and BLS1s}

Up to now, many efforts have been made to reveal the difference 
of the SEDs between NLS1s and BLS1s. 
We summarize such studies and construct template SEDs
for the NLS1s and the BLS1s which will be used 
in the following model calculations.

  \subsubsection{Observational Properties}

First, we mention the infrared properties of NLS1s and BLS1s.
Rodr\'{\i}guez-Pascual, Mas-Hesse, \& Santos-Lle\'{o} (1997) pointed out
that the FIR properties of the NLS1s and the BLS1s
are very similar to each other.
Murayama, Nagao, \& Taniguchi (1999) have reported that 
the mid-infrared properties of the NLS1s are also similar to 
those of the BLS1s.
Therefore we assume that the infrared properties of NLS1s are 
nearly the same as those of BLS1s.

Second, we mention the X-ray properties of NLS1s and BLS1s.
Boller et al. (1996) revealed that NLS1s have generally steeper 
soft X-ray spectra observed by {\it ROSAT} than BLS1s. 
The weighted mean soft X-ray photon index for their sample of NLS1s is 3.13
with an uncertainty in the mean of less than 0.03.
This is statistically larger than that of BLS1s:
the weighted mean soft X-ray photon index for the 51 BLS1s 
in the sample of Walter \& Fink (1993) is 2.34 and the uncertainty
in this mean is 0.03 (see Boller et al. 1996).
Moreover, it is known that the hard X-ray spectra of NLS1s
are also steeper than those of BLS1s.
Brandt et al. (1997) gave the average photon indices of the hard 
X-ray spectra observed by {\it ASCA} for 15 NLS1s and 19 BLS1s:
the mean hard X-ray photon index of the NLS1s is 2.15 where the
variance of this value is 0.036 and the standard error is 0.049, 
and the mean hard X-ray photon index of the BLS1s is 1.87 where the
variance of this value is 0.025 and the standard error is 0.036.

Third, we mention optical to X-ray properties of NLS1s and BLS1s.
The ratios of optical (i.e., 2500 ${\rm \AA}$) to X-ray flux at 2 keV are 
parameterized using $\alpha_{\rm ox}$, which is defined as
\begin{equation}
\alpha_{{\rm ox}} \equiv \frac{\log[F_{\nu}({\rm 2 keV})/F_{\nu}
(2500 {\rm\AA})]}{\log[\nu({\rm 2 keV})/\nu(2500 {\rm\AA})]}
\end{equation}
(Tananbaum et al. 1979).
The average $\alpha_{\rm ox}$ for optically-selected radio-quiet AGNs is
--1.46$^{+0.05}_{-0.07}$ (Zamorani et al. 1981).
The mean value of $\alpha_{\rm ox}$ derived by Puchnarewicz et al. (1996),
whose sample is X-ray selected one, is harder than the others: 
--1.14 $\pm$ 0.18.
In order to investigate whether or not this value is systematically
different between NLS1s and BLS1s, we compare $\alpha_{\rm ox}$
of 10 NLS1s\footnote{
   The NLS1s used in this test: 
   NGC 4051, NGC 4748, Mrk 359, Mrk 478, Mrk 766, I Zw 1, Akn 564,
   IRAS 13349+2438, Kaz 163, and PG 1448+273.
} and 28 BLS1s\footnote{
   The BLS1s used in this test: 
   NGC 4593, Mrk 10, Mrk 79, Mrk 142, Mrk 279, Mrk 352, Mrk 590, 
   Mrk 704, Mrk 705, Mrk 1383, 3C 263, 3C 273, 3C 382, 4C 73.18, 
   VII Zw 118, Akn 120, ES O141-G55, Fairall 9, IC 4329A, Kaz 102, 
   PG 0804+761, PG 0953+415, PG 1116+215, PG 1211+143, PG 1444+407, 
   Q 0721+343B, Q 1821+643, and Ton 1542.
} taken from the sample of Walter \& Fink (1993).
It is noted that the values of $\alpha_{\rm ox}$ for the sample of 
Walter \& Fink (1993) have slightly different from those described as
equation (1) because they measured the optical continuum
flux at 2675 ${\rm \AA}$, not at 2500 ${\rm \AA}$:
this leads to a difference of 0.02 in 
$\alpha_{\rm ox}$ (see Puchnarewicz et al. 1996).
The average spectral indices and 1$\sigma$ deviations for the 
NLS1s and the BLS1s are --1.31 $\pm$ 0.16 and --1.36 $\pm$ 
0.24, respectively.
The KS probability that the underlying distribution of these two 
distributions are the same is 5.984 $\times$ 10$^{-1}$.
Therefore there is little or no difference
in $\alpha_{\rm ox}$ between NLS1s and BLS1s.
This seems to be rather contradictory to some previous works
(Walter \& Fink 1993; Laor et al. 1994; Puchnarewicz et al. 1996), which
claimed the existence of the correlation between $\alpha_{\rm ox}$
and the X-ray spectral index, because
the large X-ray spectral index is one of the characteristic properties
of NLS1s.
The reason why such a complex situation is caused may be that
some of NLS1s in the sample of Walter \& Fink (1993) is out of
the correlation (see Figure 8 of Walter \& Fink 1993)
although it is not clear whether or not this property is a general one
in NLS1s.

  \subsubsection{SED Templates}

\begin{figure*}
\epsscale{1.4}
\plotone{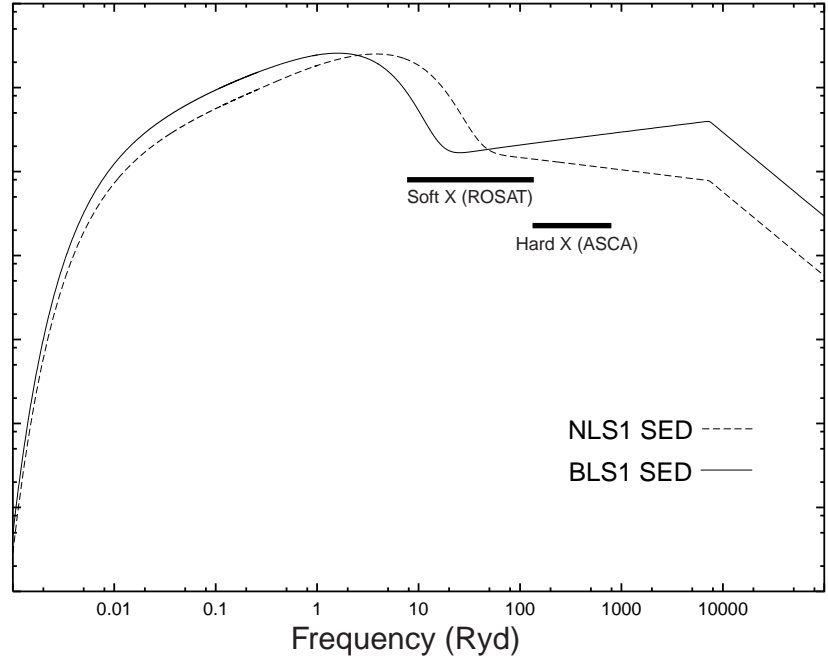}
\caption{
The template SEDs adopted for the photoionization model cluculations. 
Arbitrary flux density (i.e., energy per unit frequency) in logarithmic scale
is plotted as a function of frequency in the unit of Rydberg.
The dashed curve is for NLS1s and the solid one is for BLS1s.
\label{fig6}}
\end{figure*}

Here we construct the template SEDs of the NLS1 and the BLS1 taking the
above observational properties into account.
We adopt the following function for the templates:
\begin{equation}
f_{\nu} = \nu^{\alpha_{{\rm uv}}} \exp(-\frac{h\nu}{kT_{{\rm BB}}}) \exp
(-\frac{kT_{{\rm IR}}}{h\nu}) + a\nu^{\alpha_{{\rm x}}}
\end{equation}
(see Ferland 1996).
We adopt the following parameter values.
(I) $kT_{\rm IR}$ is the infrared cutoff of the so-called big blue bump 
component, and we fix $kT_{\rm IR}$ = 0.01 Ryd following to Ferland (1996).
(II) $\alpha_{\rm UV}$ is the slope of the low-energy side of the big blue
bump component. We adopt $\alpha_{\rm UV}$ = --0.5, which is the typical value
for AGNs (Ferland 1996; see also Francis 1993).
Note that the photoionization process is not sensitive to this parameter.
(III) $\alpha_{\rm ox}$ is the UV--to--X-ray spectral slope mentioned above,
which determines the parameter $a$ in the equation (2).
We adopt $\alpha_{\rm ox}$ = --1.35, which is the average value for the sample
of Walter \& Fink (1993) mentioned above, for both sample.
However, there are some claims that this parameter correlates to the
X-ray spectral index (Walter \& Fink 1993; Laor et al. 1994 
Puchnarewicz et al. 1996),
that is, $\alpha_{\rm ox}$ may be different between NLS1s and BLS1s.
Hence we check the dependence of the calculation outputs on 
$\alpha_{\rm ox}$ in Section 4.3.3.
(IV) $\alpha_{\rm x}$ is the slope of the X-ray component.
We adopt $\alpha_{\rm x}$ = --1.15 for NLS1s 
and --0.85 for BLS1s corresponding to
the observational results by $ASCA$ described in Section 4.1.1.
This power-law component is not extrapolated below 1.36 eV or above 100 keV.
Below 1.36 eV, this term is set to zero, while above 100 keV, the 
continuum is assumed to fall off as $\nu^{-3}$.
And finally, (V) $T_{{\rm BB}}$ is the parameter which characterizes the 
shape of the
big blue bump. We choose this parameter to reproduce the soft X-ray 
spectral index measured by {\it ROSAT} described in Section 4.1.1. 
It results in 1,180,000 K for NLS1s and
490,000 K for BLS1s. They correspond to $\Gamma_{\rm ROSAT}$ = 3.13 and 
2.35, respectively.
The template SEDs constructed in such a way are shown in Figure 6;
hereafter we refer the NLS1 SED and the BLS1 SED, respectively.

It is notable that these template SEDs are not theoretically predicted ones,
but the empirical ones. 
Although it has not been understood whether or not the soft excess component 
is well described by a blackbody, Pounds et al. (1994) mentioned that the 
soft excess can be characterized by a blackbody of 
$kT_{\rm BB}$ = 70 $\pm$ 10 eV [or $T_{\rm BB} = (8.1 \pm 1.2) \times 10^5$ K].
This suggests that the temperature of our adopted SEDs is not too high.
Puchnarewicz et al. (1996) also mentioned that the soft excess may be
represented by thermal bremsstrahlung with $T_{\rm brem}$ = 10$^6$ K.
Mineshige et al. (2000) proposed a slim disk model whose maximum blackbody 
temperature is $kT_{\rm BB} \simeq 0.2 (M_{\rm SMBH}/10^5 M_{\odot})^{-1/4}$ 
keV [or $T_{\rm BB} \simeq 2.3 \times 10^6 
(M_{\rm SMBH}/10^5 M_{\odot})^{-1/4}$]
for the soft excess of NLS1s where $M_{\rm SMBH}$ is the mass
of a supermassive black hole.
These studies are almost consistent with our empirical template SEDs.

 \subsection{The Calculation Procedure}

We perform photoionization model calculations using the 
spectral synthesis code {\it Cloudy} version 90.04 (Ferland 1996), 
which solves the 
equations of statistical and thermal equilibrium and produces a 
self-consistent model of the run of temperature as a function of depth into 
the nebula. Here we assume an uniform density gas cloud with 
a plane-parallel geometry.

The parameters for the calculations are (1) the hydrogen density of 
the cloud ($n_{\rm H}$), (2) the ionization parameter ($U$), which is defined 
as the ratio of the ionizing photon density to the electron density,
(3) the chemical compositions of the gas, and (4) the shape of the input SED.
We perform several model runs covering the following ranges of parameters:
10$^3$ cm$^{-3} \leq n_{\rm H} \leq 10^6$ cm$^{-3}$ (and 10$^7$ cm$^{-3}$
in Section 4.3.4) and 10$^{-4} \leq U \leq 10^{-1}$.
We set the gas-phase elemental abundance to be either solar or subsolar.
The adopted solar abundances relative to hydrogen are taken from
Grevesse \& Anders (1989) with extensions by Grevesse \& Noels (1993).
The subsolar abundances are assumed that 90\% of Mg, Si, and Fe,
50\% of C and O, and 25\% of N and S are locked into dust grains,
as estimated for the Orion H {\sc ii} region (e.g., Baldwin et al. 1991, 1996).
For the input SEDs, we use the two types of SED: the NLS1 SED and 
the BLS1 SED, mentioned in the last section.
The calculations are stopped when the temperature fall to 3000 K, below which
gas does not contribute significantly to the optical emission lines.

 \subsection{The Results of the Calculations}

  \subsubsection{Excitation}

\begin{figure*}
\epsscale{1.0}
\plotone{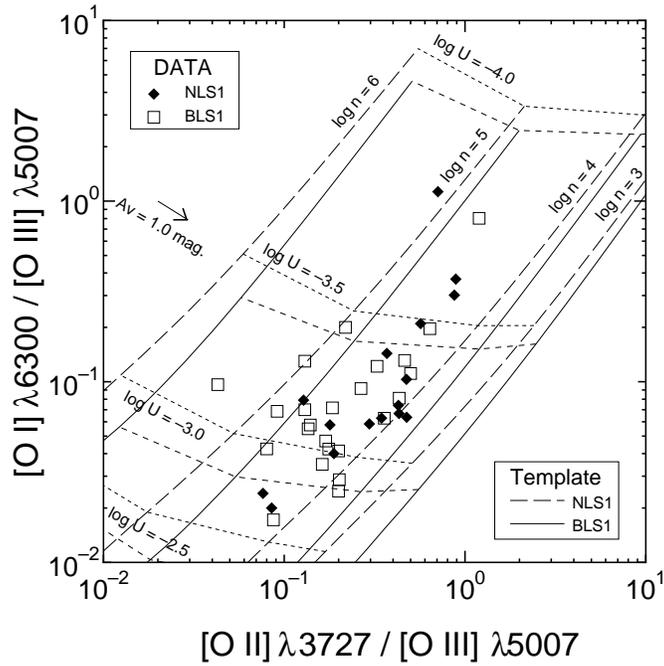}
\caption{
The diagram of [O {\sc i}]/[O {\sc iii}]$\lambda$5007 versus
[O {\sc ii}]/[O {\sc iii}]$\lambda$5007. 
The symbols are the same as in Figure 3.
Our model calculations for the solar abundances are presented by dashed curves
for the NLS1 SED and by solid curve for the BLS1 SED.
The data points will move on the diagram as shown by the arrow 
if the extinction correlation of $A_V$ = 1.0 is applied.
\label{fig7}}
\end{figure*}

\begin{figure*}
\epsscale{1.0}
\plotone{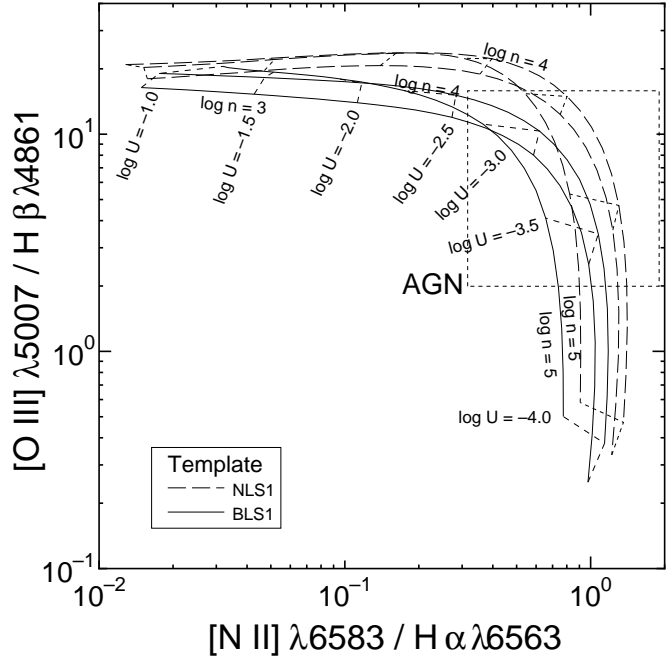}
\caption{
The diagram of [O {\sc iii}]$\lambda$5007/H$\beta$ versus 
[N {\sc ii}]/H$\alpha$.
The model loci for the case of the solar abundances are 
the same as in Figure 7.
Typical location of AGNs (see Figure 4 of Veilleux \& Osterbrock 1987)
on this diagram are shown by the box.
The effect of the correction for the extinction is not shown in this figure 
because the reddening effect on these line ratios is negligiblly small.
\label{fig8}}
\end{figure*}

\begin{figure*}
\epsscale{1.0}
\plotone{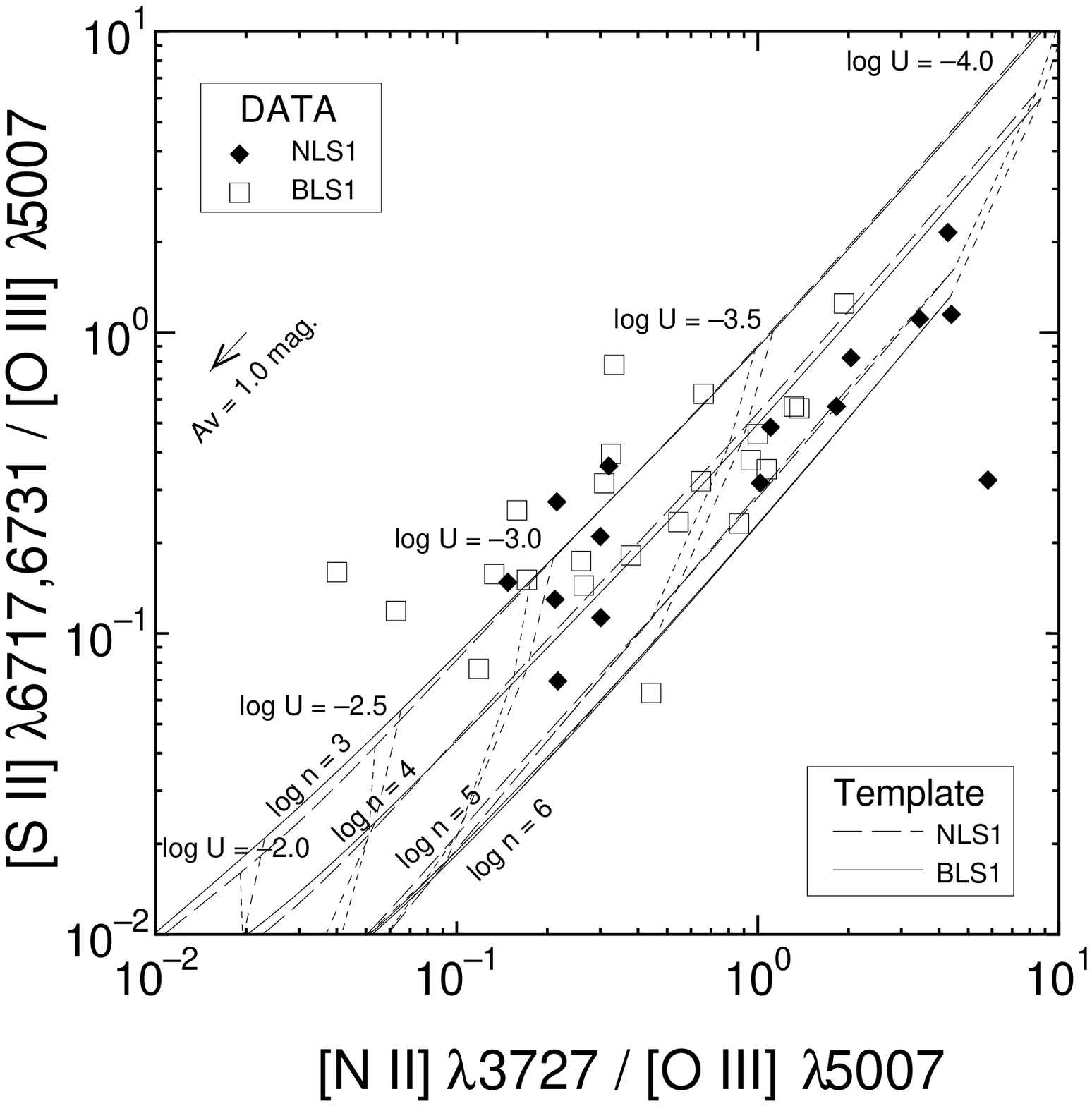}
\caption{
The diagram of [S {\sc ii}]/[O {\sc iii}]$\lambda$5007 versus
[N {\sc ii}]/[O {\sc iii}]$\lambda$5007.
The results of the model calculations for the case of the solar abundance 
are shown.
The lines, the symbols, and the arrow are the same as in Figure 7.
\label{fig9}}
\end{figure*}

\begin{figure*}
\epsscale{1.0}
\plotone{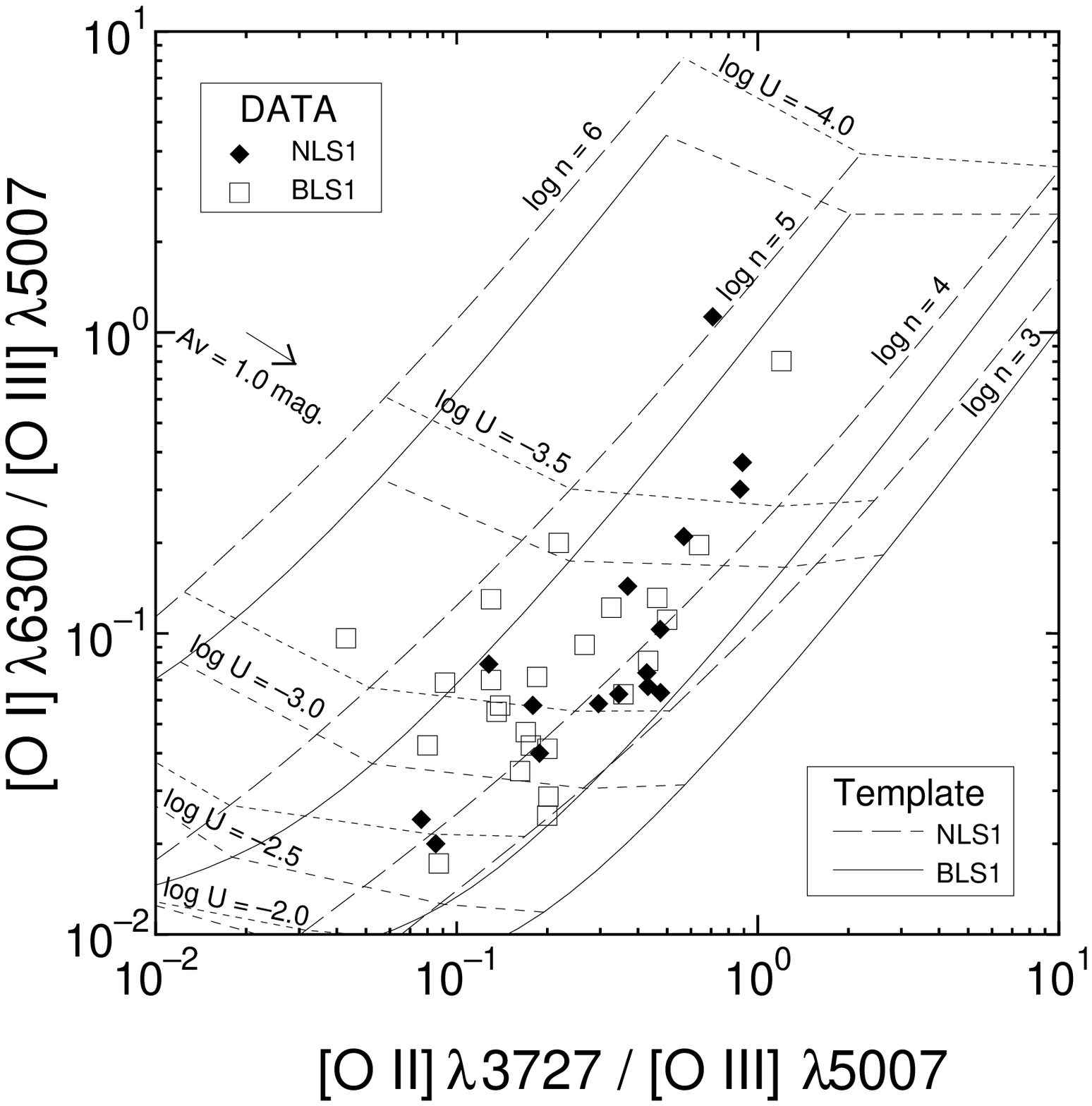}
\caption{
The diagram of [O {\sc i}]/[O {\sc iii}]$\lambda$5007 versus
[O {\sc ii}]/[O {\sc iii}]$\lambda$5007.
The results of the model calculations for the case of the subsolar abundance 
are shown.
The lines, the symbols, and the arrow are the same as in Figure 7.
\label{fig10}}
\end{figure*}

\begin{figure*}
\epsscale{1.0}
\plotone{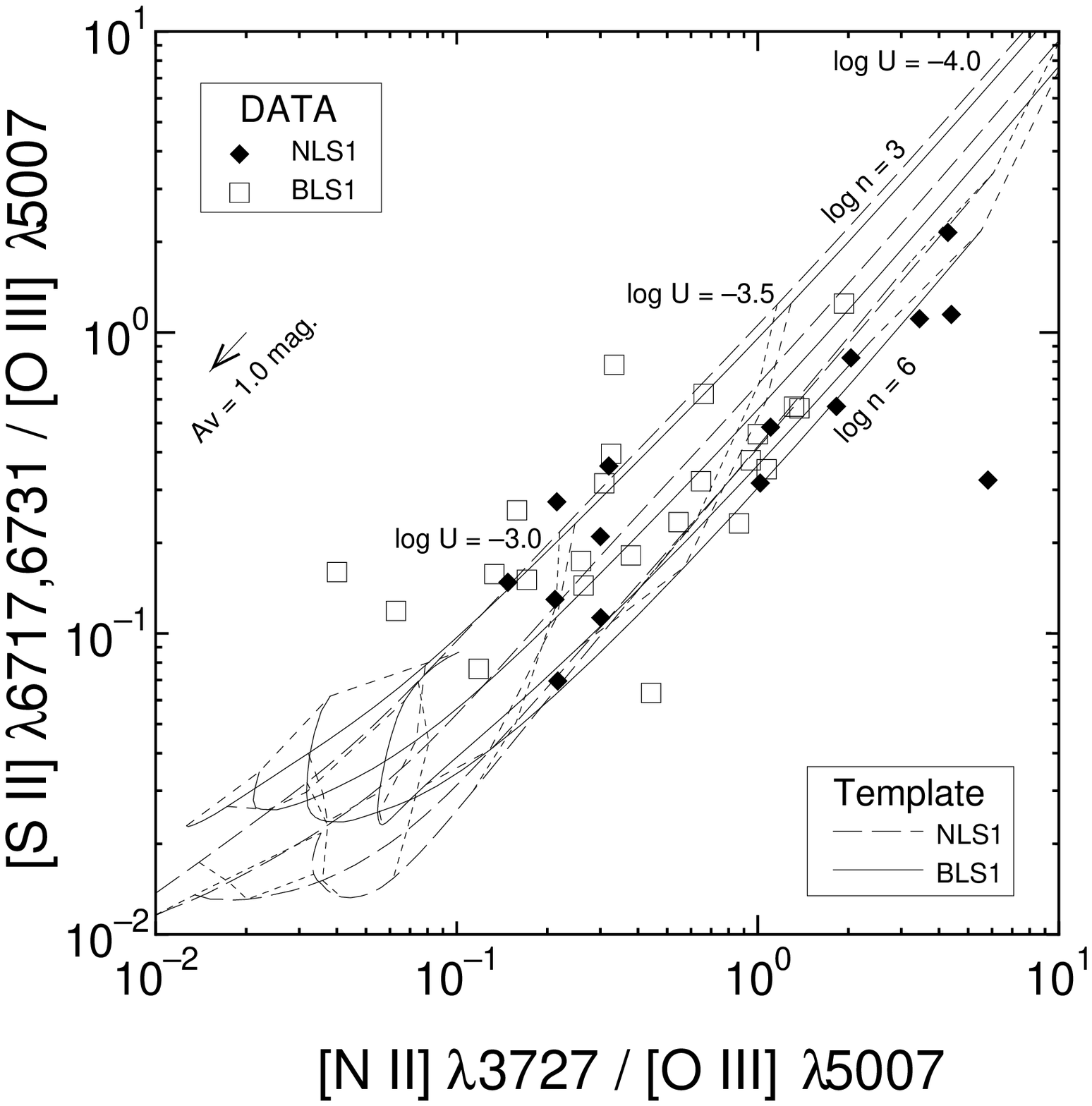}
\caption{
The diagram of [S {\sc ii}]/[O {\sc iii}]$\lambda$5007 versus
[N {\sc ii}]/[O {\sc iii}]$\lambda$5007.
The results of the model calculations for the case of the subsolar abundance 
are shown.
The lines, the symbols, and the arrow are the same as in Figure 7.
\label{fig11}}
\end{figure*}

We show the results of the model calculations in the case of the solar 
abundances and compare them to the observations
in Figure 7, which is a diagram of [O {\sc i}]/[O {\sc iii}]$\lambda$5007
versus [O {\sc ii}]/[O {\sc iii}]$\lambda$5007.
This diagram has been used to discuss the physical properties of 
ionized gas traditionally (e.g., Heckman 1980; Baldwin, Phillips, \& Terlevich
1981; Ferland \& Netzer 1983; Evans et al. 1999).
It is shown that there is a slight difference of 
[O {\sc i}]/[O {\sc iii}]$\lambda$5007 between the two model grids.
The reason for this difference is thought as follows.
The relative intensity of soft X-ray of NLS1s is stronger than that of
BLS1s. This results in a larger partially ionized zone
in NLRs of NLS1s. 
Since [O {\sc i}] is selectively radiated from such partially ionized 
zone because its ionization potential is close to the ionization potential 
of hydrogen, NLS1s tend to exhibit stronger [O {\sc i}].
However, the dispersion of the compiled data is larger than this difference.
This means that {\it the line ratios used in this diagram
are insensitive to the difference of the shape of the template SEDs}.
This result is consistent with the previous work of Rodr\'{\i}guez-Ardila 
et al. (2000a). They presented their photoionization
model calculations assuming two types of SEDs; i.e., the NLS1-like SED and 
the BLS1-like SED\footnote{
   The shapes of their template SEDs are simpler than ours.
   They are double power-law functions, whose powers are based on the
   spectral indices found from ROSAT and ASCA data.
   See Rodr\'{\i}guez-Ardila et al. (2000a) for the details of their models.
}. As shown in Figure 7 of Rodr\'{\i}guez-Ardila et al. 
(2000a), the ratio of [O {\sc i}]/[O {\sc iii}]$\lambda$5007 is not so 
different between the model NLS1 and the model BLS1 (less than factor 3).

The comparison between the models and the observations shown in Figure 7
suggests $10^4 {\rm cm}^{-3} \leq n_{\rm H} \leq 10^5 {\rm cm}^{-3}$ and 
$10^{-3.5} \leq U \leq 10^{-3}$ for the NLRs of both samples.
The estimated $U$ values seem to be rather lower than those calculated in
some previous works (e.g., Ferland \& Netzer 1983; 
Ho, Shields, \& Filippenko 1993).
In order to make it clear that this is not due to any selection effect 
of the samples, we show model grids on the diagram of 
[O {\sc iii}]$\lambda$5007/H$\beta$ versus [N {\sc ii}]/H$\alpha$, 
which is a familiar diagnostic diagram proposed by 
Veilleux \& Osterbrock (1987), in Figure 8.
Comparing this with Figure 4 of Veilleux \& Osterbrock (1987), 
typical Seyferts are reproduced by the models with
$10^{-3.5} \leq U \leq 10^{-3}$.
This discrepancy between the result by us and by previous literature
is thought to be partly because the following reason.
The energy peak of the template SEDs in our models is at rather high energy
than those in previous studies. Accordingly the relative amounts of photons
whose energy is near the ionization potential of hydrogen increase.
This leads to the lower $U$ because this parameter is defined using 
{\it all} photons which exceed the ionization potential of hydrogen 
although the photoionization is effective in the energy of near the 
ionization potential of hydrogen.

We investigate the gas properties with another diagnostic diagram:
[S {\sc ii}]/[O {\sc iii}]$\lambda$5007 versus
[N {\sc ii}]/[O {\sc iii}]$\lambda$5007 (Figure 9).
It results in that
the derived ranges in both $n_{\rm H}$ and $U$ are consistent 
with those obtained in Figure 7.
There is very little difference between the model grids for NLS1s 
and those for BLS1s.
It is notable that the scatter of the plotted data in this diagram is
larger than that in Figure 6. This may be due to that
the deblending H$\alpha$ from [N {\sc ii}] is not be well done
in some case if the spectral resolution is not so high.
If this is the case, the flux measurement of H$\alpha$ may not also
be well done. It means that it is dangerous to use traditional emission-line 
ratios such as [N {\sc ii}]/H$\alpha$, [S {\sc ii}]/H$\alpha$, and 
[O {\sc i}]/H$\alpha$ for S1s.
Alternatively, the scatter may reflect the variety of the nitrogen 
abundance because some previous works reported that some of Seyferts show
evidences in favor of a nitrogen overabundance (Storchi-Bergmann \&
Pastoriza 1990; Storchi-Bergmann 1991; Storchi-Bergmann et al. 1998).
In any case, the diagram of [S {\sc ii}]/[O {\sc iii}]$\lambda$5007 versus
[N {\sc ii}]/[O {\sc iii}]$\lambda$5007 is less suitable to discuss the
properties of gas in NLRs than that of
[O {\sc i}]/[O {\sc iii}]$\lambda$5007
versus [O {\sc ii}]/[O {\sc iii}]$\lambda$5007.

In Figures 10 and 11, we show the result of the model calculations 
for the case of the subsolar abundances and compare them with the observations
on the diagrams of [O {\sc i}]/[O {\sc iii}]$\lambda$5007
versus [O {\sc ii}]/[O {\sc iii}]$\lambda$5007 and
[S {\sc ii}]/[O {\sc iii}]$\lambda$5007 versus
[N {\sc ii}]/[O {\sc iii}]$\lambda$5007, respectively.
The loci of model grids in Figure 10 slightly shift to be larger
in [O {\sc i}] than
those in Figure 7. This may be attributed to the fact that
the partially ionized region become thicker due to the decrease of the 
heavy elements.
However, the estimated parameters, $n_{\rm H}$ and $U$, 
are almost the same as those in the case of the solar abundances.

  \subsubsection{[O {\sc iii}] Emitting Region}

\begin{figure*}
\epsscale{1.0}
\plotone{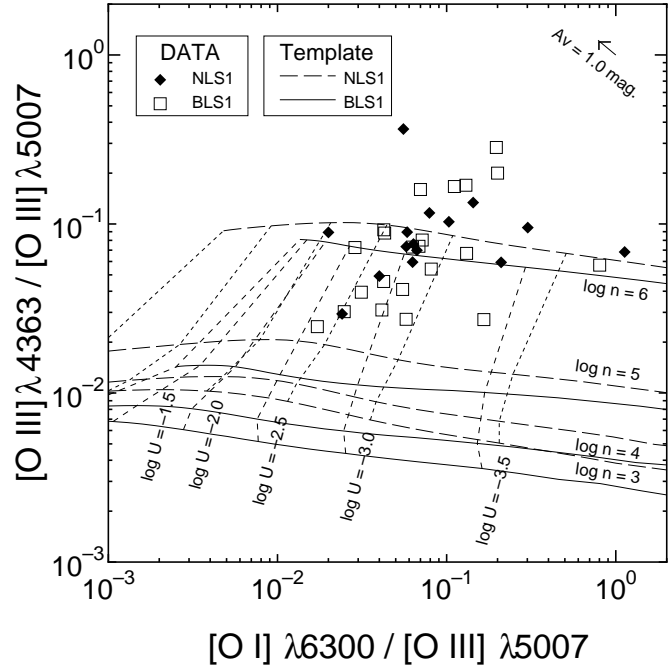}
\caption{
The diagram of [O {\sc iii}]$\lambda$4363/[O {\sc iii}]$\lambda$5007 versus
[O {\sc i}]/[O {\sc iii}]$\lambda$5007.
The results of the model calculations for the case of the solar abundance 
are shown.
The lines, the symbols, and the arrow are the same as in Figure 7.
\label{fig12}}
\end{figure*}

\begin{figure*}
\epsscale{1.0}
\plotone{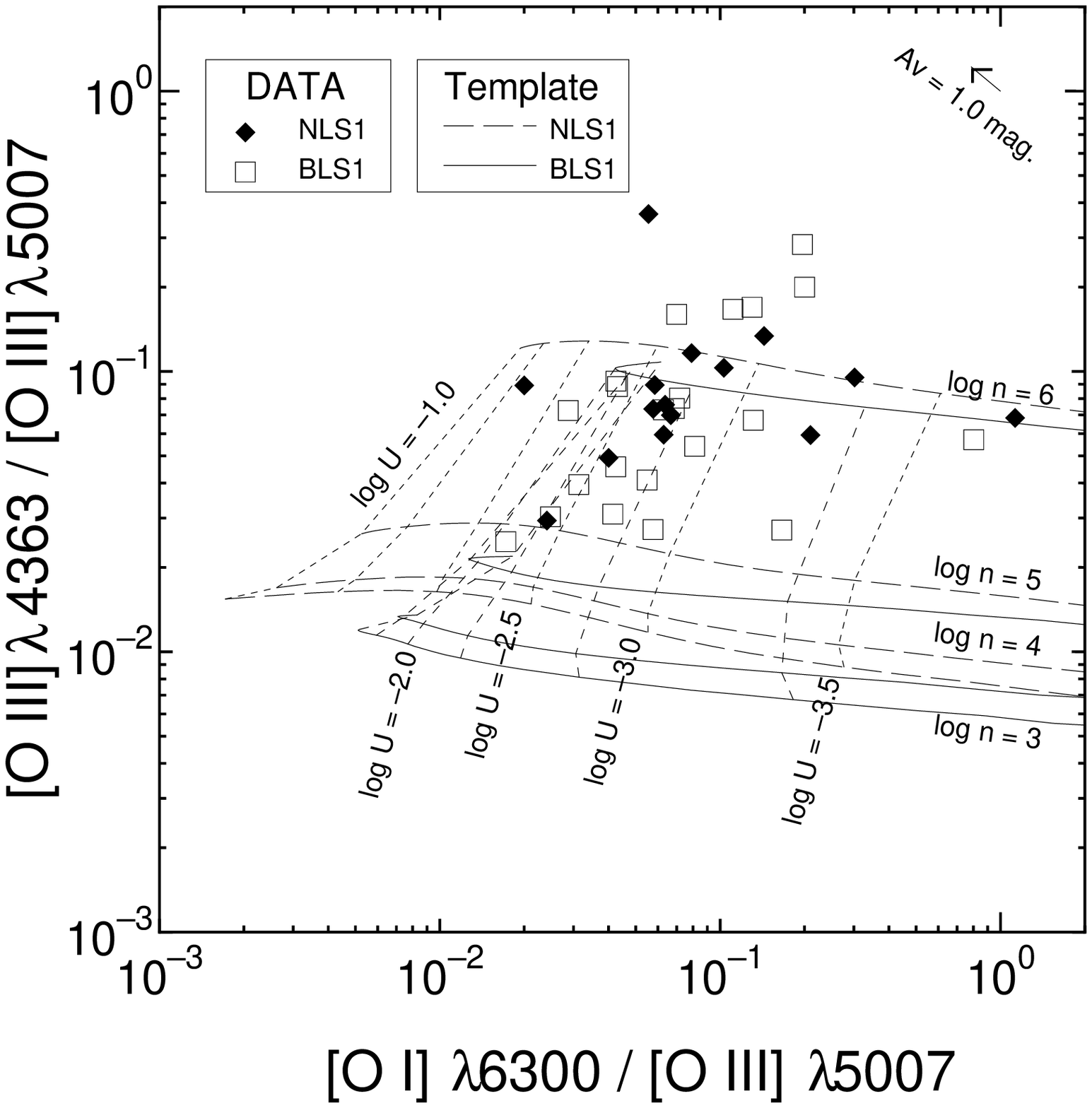}
\caption{
The diagram of [O {\sc iii}]$\lambda$4363/[O {\sc iii}]$\lambda$5007 versus
[O {\sc i}]/[O {\sc iii}]$\lambda$5007.
The results of the model calculations for the case of the subsolar abundance 
are shown.
The lines, the symbols, and the arrow are the same as in Figure 7.
\label{fig13}}
\end{figure*}

\begin{figure*}
\epsscale{1.0}
\plotone{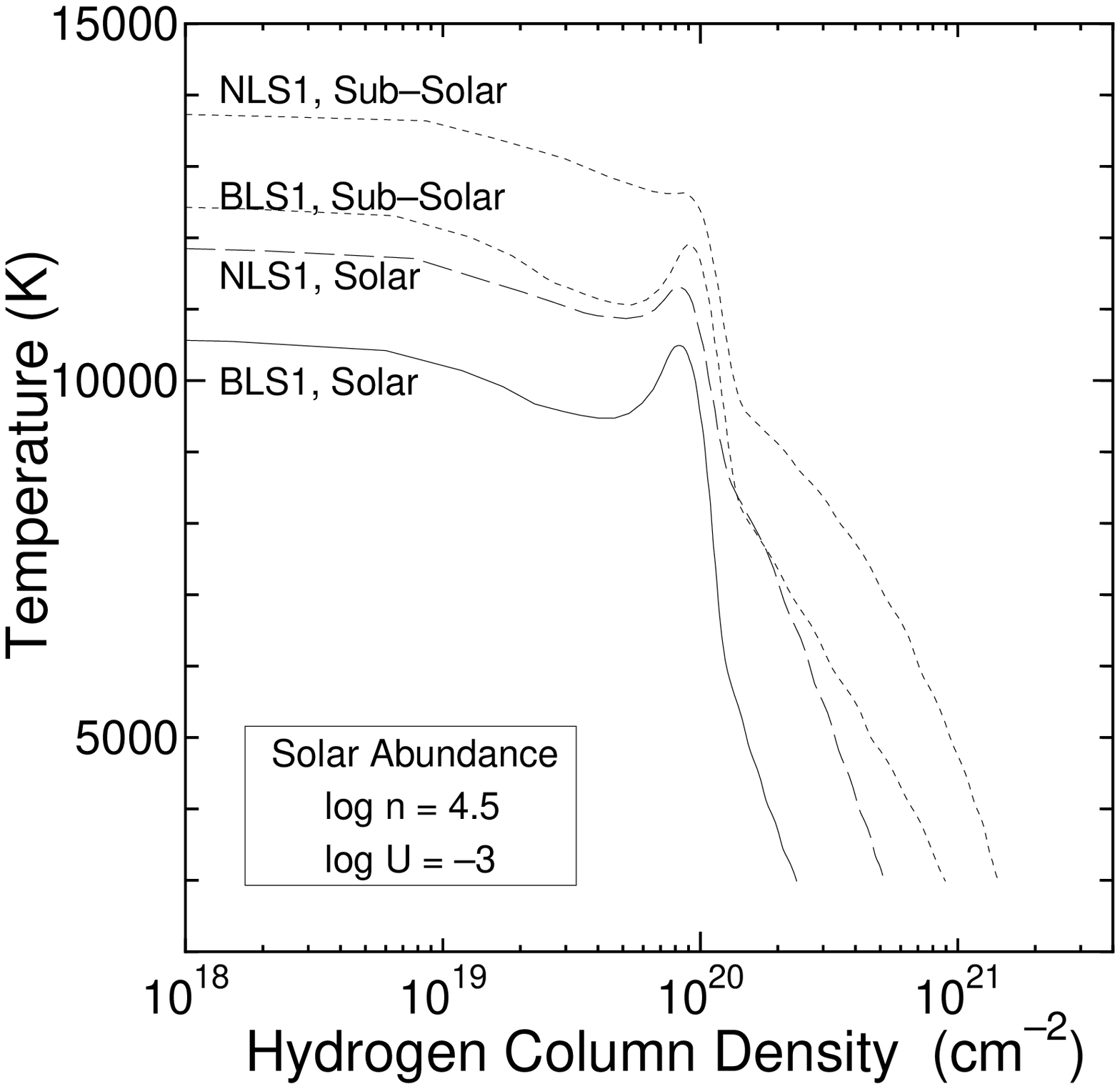}
\caption{
The gas temperature calculated with the models described in the text
is shown
as a function of the hydrogen column density
from the surface of a cloud exposed to the ionizing source.
\label{fig14}}
\end{figure*}

As mentioned in Section 3.2, the temperature-sensitive emission-line ratio,
[O {\sc iii}]$\lambda$4363/[O {\sc iii}]$\lambda$5007, is scarcely 
different between the NLS1s and the BLS1s though the intrinsic SEDs
are clearly different between them.
In order to investigate whether or not this difference in SEDs causes
a detectable difference in the temperature of NLR gas through 
photoionization processes,
we carry out the model calculations concerning
the [O {\sc iii}]$\lambda$4363/[O {\sc iii}]$\lambda$5007 ratio.
In Figure 12, we show the diagram of
[O {\sc iii}]$\lambda$4363/[O {\sc iii}]$\lambda$5007
versus [O {\sc i}]/[O {\sc iii}]$\lambda$5007 for the case of 
the solar abundances.
It is shown that the very high density condition
(10$^6 {\rm cm}^{-3}$ or higher) is needed
to explain the observed [O {\sc iii}]$\lambda$4363/[O {\sc iii}]$\lambda$5007
ratios for both NLS1s and BLS1s.
This derived $n_{\rm H}$ is far higher than the value obtained using
the diagnostic diagrams presented in Section 4.3.1.
The reason for this is partly
because of the oversimplification of the photoionization models:
the [O {\sc iii}]$\lambda$4363/[O {\sc iii}]$\lambda$5007 ratio
is difficult to be reproduced by one-zone 
photoionization models as mentioned in other works
(e.g., Filippenko \& Halpern 1984; Tadhunter, Robinson, \& Morganti 1989;
Wilson, Binette, \& Storchi-Bergmann 1997;
Nagao, Murayama, \& Taniguchi 2000a).
We are not going to make further discussion for this problem 
because this is out of the purpose of this paper.
In this diagram, the loci of the model grid for NLS1s slightly shift
to be larger [O {\sc iii}]$\lambda$4363/[O {\sc iii}]$\lambda$5007
with respect to those for BLS1s,
which means that the temperature of gas in NLRs of the model NLS1 
is higher than that of the model BLS1. 
This is because the number of high energy photon
is larger in NLS1s than in BLS1s (see Figure 6).
However, it is evident that this difference of the model loci is
much smaller than the dispersion of the observed data points.
Therefore, we conclude that the difference of SEDs between the NLS1s and 
the BLS1s is not important when one investigates the 
[O {\sc iii}]$\lambda$4363/[O {\sc iii}]$\lambda$5007 ratio.
As shown in Figure 13, almost the same results are obtained
when the subsolar abundances are assumed on the model calculations.

In order to investigate the difference of gas temperature between NLS1s
and BLS1s in more detail, we show the temperature structure
in a cloud as a function of
the hydrogen column density from the inner surface in Figure 14.
Here we adopt $n_{\rm H} = 10^{4.5}$ cm$^{-3}$ and $U = 10^{-3}$
for the four model calculations.
It is shown that the gas temperature in the case of 
the subsolar abundances is higher than the other, which is due to
a decrease of coolant elements.
Generally the temperature decreases as the column density increases.
However, there is a small turn-up just before the ionization front
where the temperature drops off.
This is attributed to the fact that the most effective coolant, 
O$^{2+}$, is exhausted at this region.
It is important that the abundances more affect the temperature
than the input SED.
Therefore, the difference of SEDs between the NLS1s and BLS1s
does not affect significantly the gas temperature proved by the 
[O {\sc iii}] emission lines.
This result is also consistent with Rodr\'{\i}guez-Ardila et al. (2000a),
in which the ratio of [O {\sc iii}]$\lambda$4363/[O {\sc iii}]$\lambda$5007 
is not so different between  the model NLS1 and the model BLS1 
(less than factor 3) though they suggested that the calculated ratio of
[O {\sc iii}]$\lambda$4363/[O {\sc iii}]$\lambda$5007 is higher in the BLS1
than in the NLS1.

  \subsubsection{Dependence of the Calculation Results on $\alpha_{\rm ox}$}

\begin{figure*}
\epsscale{1.0}
\plotone{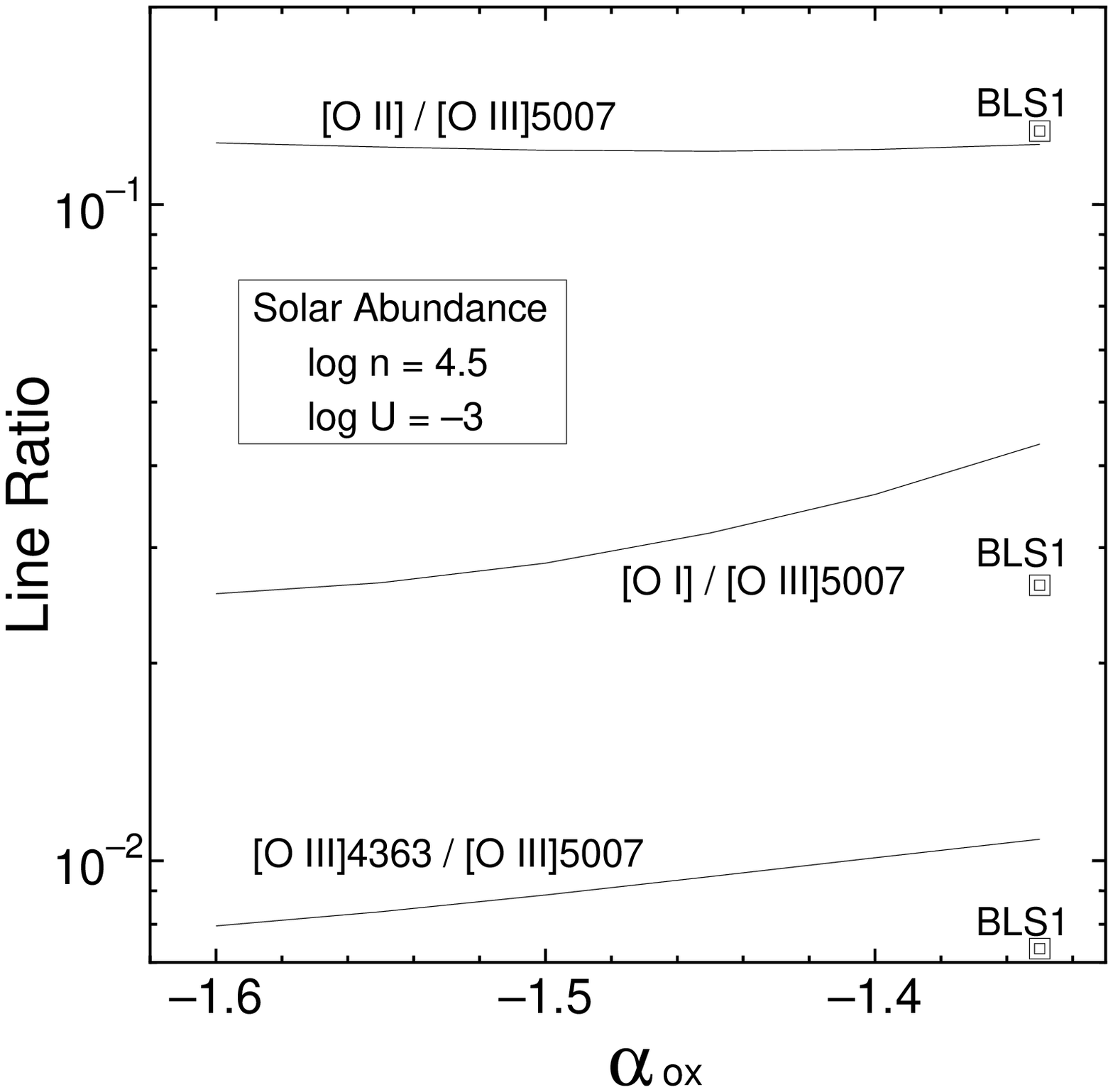}
\caption{
The calculated line ratios of [O {\sc i}]/[O {\sc iii}]$\lambda$5007,
[O {\sc ii}]/[O {\sc iii}]$\lambda$5007, and
[O {\sc iii}]$\lambda$4363/[O {\sc iii}]$\lambda$5007 versus $\alpha_{\rm ox}$
for the NLS1 SED.
The calculated line ratios for the BLS1 SED ($\alpha_{\rm ox}$ = --1.35)
are plotted by squares.
For the calculations, $n_{\rm H} = 10^{4.5}$ cm$^{-3}$, $U = 10^{-3}$ and
solar abundances are assumed.
\label{fig15}}
\end{figure*}

\begin{figure*}
\epsscale{1.0}
\plotone{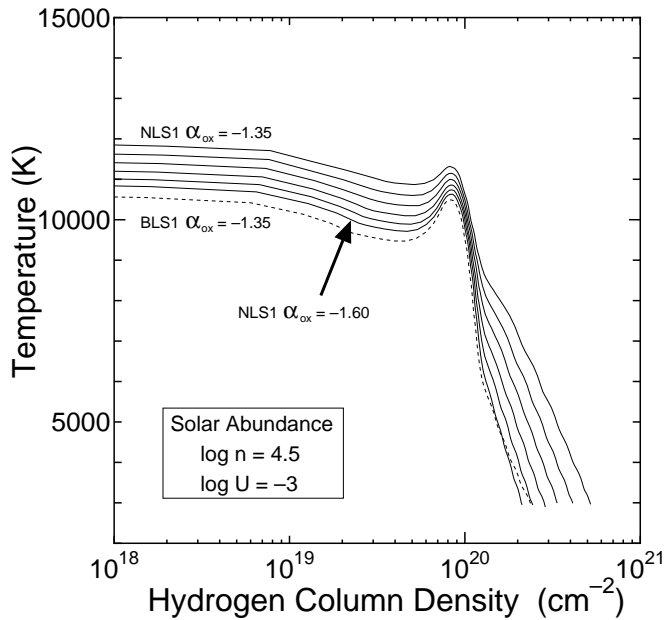}
\caption{
The calculated gas temperature of a NLR gas of NLS1s 
for various $\alpha_{\rm ox}$ are shown by solid lines
for the case of the NLS1 SED.
The case of BLS1 SED ($\alpha_{\rm ox}$ = --1.35) is shown by dotted curve.
For the calculations, $n_{\rm H} = 10^{4.5}$ cm$^{-3}$, $U = 10^{-3}$ and
solar abundances are assumed.
\label{fig16}}
\end{figure*}

As mentioned in Section 4.1, we have assumed that NLS1s and
BLS1s have similar $\alpha_{\rm ox}$.
However, there are some previous studies (Walter \& Fink 1993; 
Laor et al. 1994; Puchnarewicz et al. 1996)
in which it is claimed that the soft X-ray spectral index 
correlates with $\alpha_{\rm ox}$.
Since the large X-ray spectral index is one of the characteristic properties
of NLS1s, their claim means that NLS1s have the softer $\alpha_{\rm ox}$
than BLS1s, systematically.
Therefore, we investigate the dependence of the calculations 
on $\alpha_{\rm ox}$.

When various values of $\alpha_{\rm ox}$ are adopted, $T_{\rm BB}$ must
be correspondingly adjusted to reproduce the observed soft X-ray photon index,
$\Gamma_{\rm ROSAT}$ = 3.13.
We adopt $T_{\rm BB}$ = 980,000 K, 840,000 K, 730,000 K, 650,000 K,
and 590,000 K for the cases of $\alpha_{\rm ox}$ = --1.40, --1.45,
--1.50, --1.55, and --1.60, respectively.

In Figure 15, we show a diagram of calculated line ratios 
versus $\alpha_{\rm ox}$, adopting $n_{\rm H} = 10^{4.5}$ cm$^{-3}$, 
$U = 10^{-3}$, solar abundances, and the SED template of NLS1s.
It is clearly shown that the [O II]/[O III]$\lambda$5007 ratio
is almost independent of $\alpha_{\rm ox}$.
On the other hand, the [O I]/[O III]$\lambda$5007 and the
[O III]$\lambda$4363/[O III]$\lambda$5007 ratios 
become smaller and to be close to
the value of BLS1s as $\alpha_{\rm ox}$ becomes softer.
Thus, we conclude that
the difference of intrinsic SEDs between NLS1s and BLS1s scarcely
affects the NLR emission even if $\alpha_{\rm ox}$ of NLS1s is systematically
softer than that of BLS1s.

In Figure 16, we show the temperature structure in a NLR cloud for various
values of $\alpha_{\rm ox}$, 
adopting $n_{\rm H} = 10^{4.5}$ cm$^{-3}$, $U = 10^{-3}$,
solar abundances, and the SED template of NLS1s.
The gas temperature also becomes to be close to that of BLS1s 
as $\alpha_{\rm ox}$ becomes softer.

  \subsubsection{Highly Ionized Emission Lines}

\begin{figure*}
\epsscale{1.0}
\plotone{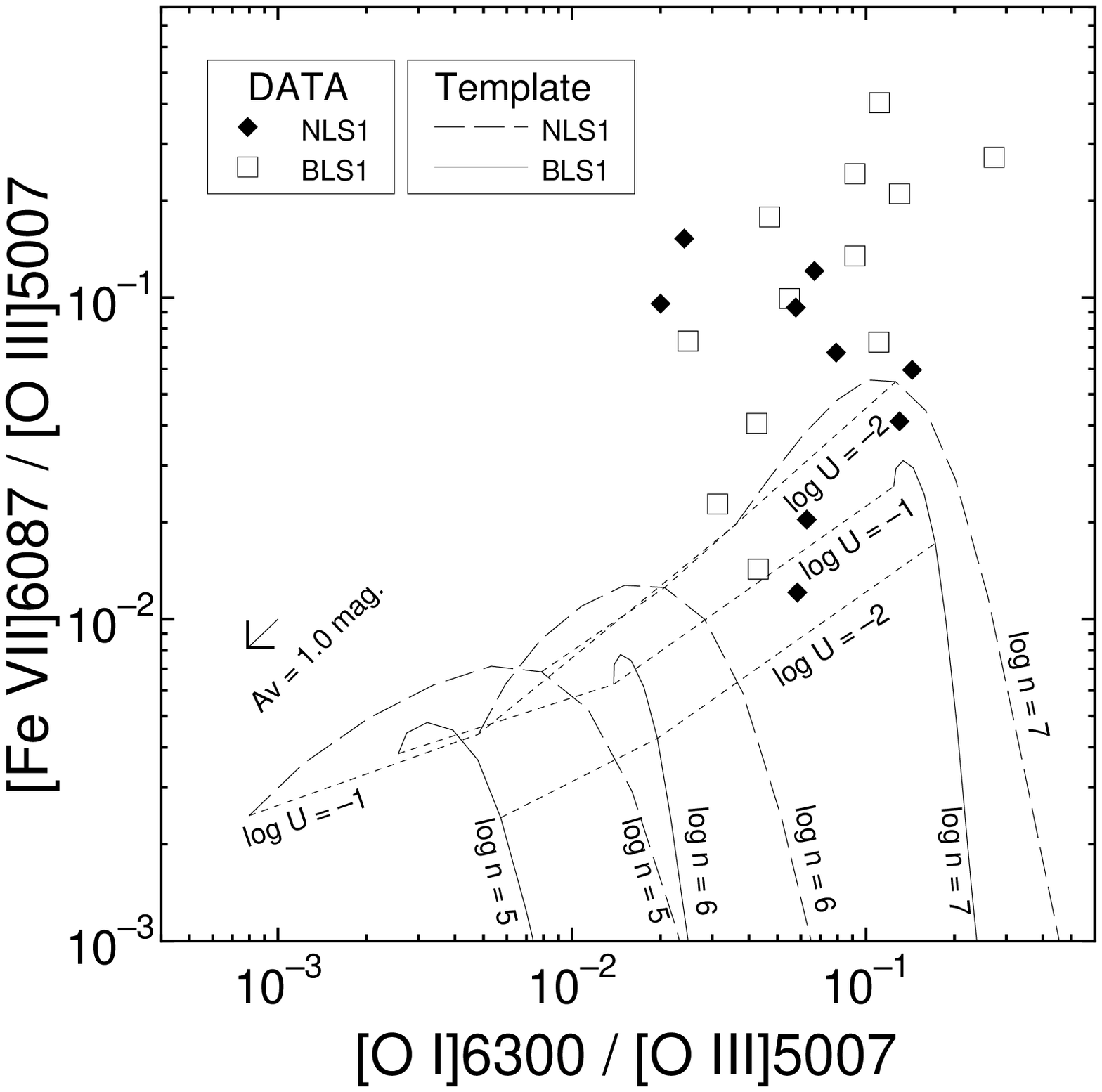}
\caption{
The diagram of [Fe {\sc vii}]$\lambda$6087/[O {\sc iii}]$\lambda$5007 versus
[O {\sc i}]/[O {\sc iii}]$\lambda$5007.
The results of the model calculations 
for the case of solar abundances are shown.
The lines, the symbols, and the arrow are te same as in Figure 7.
\label{fig17}}
\end{figure*}

\begin{figure*}
\epsscale{1.0}
\plotone{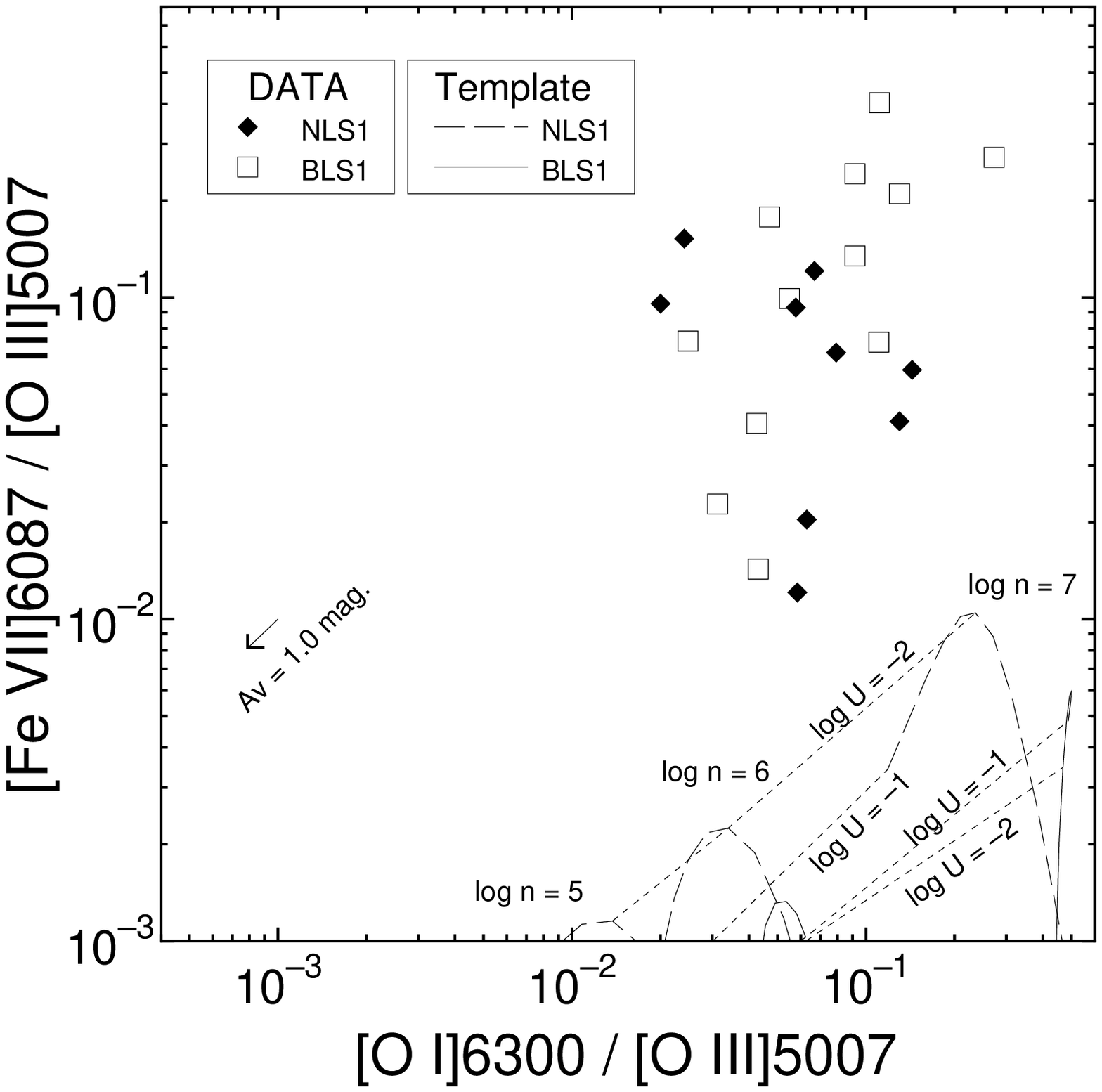}
\caption{
The diagram of [Fe {\sc vii}]$\lambda$6087/[O {\sc iii}]$\lambda$5007 versus
[O {\sc i}]/[O {\sc iii}]$\lambda$5007.
The results of the model calculations 
for the case of subsolar abundances are shown.
The lines, the symbols, and the arrow are te same as in Figure 7.
\label{fig18}}
\end{figure*}

Seyfert galaxies often present highly ionized emission lines such as
[Fe {\sc vii}]$\lambda$6087, [Fe {\sc x}]$\lambda$6374, 
[Fe {\sc xi}]$\lambda$7892, and [Fe {\sc xiv}]$\lambda$5303
(see Nagao et al. 2000c and references therein).
These emission lines are useful to investigate 
the viewing angle toward dusty tori
of Seyfert nuclei (Murayama \& Taniguchi 1998a; Nagao et al. 2000c).
Therefore it is important to investigate how the feature of intrinsic SEDs 
affects such highly ionized emission-line intensities.

We show the results of the model calculations for the case of 
the solar abundances and compare them with the observations in 
a diagram of [Fe {\sc vii}]$\lambda$6087/[O {\sc iii}]$\lambda$5007
versus [O {\sc i}]/[O {\sc iii}]$\lambda$5007 (Figure 17). 
The data of observations are taken from Nagao et al. (2000c).
Being different from the results described in Section 4.3.1 and 4.3.2, 
there are 
evident differences in the behavior of the calculated line ratios between
two models as follows. (1) The calculated 
[O {\sc i}]/[O {\sc iii}]$\lambda$5007 ratio
for NLS1s is smaller than that for BLS1s
when $U \sim 10^{-1}$ although the opposite trend is seen when $U < 10^{-2}$.
This is because the volume of the fully ionized region becomes larger
with increasing ionization parameter,
and thus the [O {\sc iii}]$\lambda$5007 emission
becomes more prominent relative to the [O {\sc i}] emission. 
(2) The calculated 
[Fe {\sc vii}]$\lambda$6087/[O {\sc iii}]$\lambda$5007 ratio for NLS1s is 
several times larger than that for BLS1s when $U \leq 10^{-2}$.
This is because the number of the high-energy ionizing photons\footnote{
   The ionization potential of the lower stage of ionization 
   and the critical density for [Fe {\sc vii}]
   is 99.1 eV and 3.6 $\times 10^7$ cm$^{-3}$, respectively.
} producing Fe$^{6+}$ in the model
for NLS1s is much larger than that in the model for BLS1s when we adopt 
the same ionization parameter and the gas density for both cases.

However, clearly shown in Figure 17, the calculated 
[Fe {\sc vii}]$\lambda$6087/[O {\sc iii}]$\lambda$5007 is much smaller than
the observed one in both models.
This means that another component which radiates highly ionized emission 
lines is needed to explain the observations, which is consistent with 
previous studies (Stasi\'{n}ska 1984; Ferland \& Osterbrock 1986; 
Binette et al. 1996;
Murayama \& Taniguchi 1998a, 1998b).
Therefore this result {\it does not} suggest that the observed
[Fe {\sc vii}]$\lambda$6087/[O {\sc iii}]$\lambda$5007 of NLS1s should be
larger than that of BLS1s.

In Figure 18, we show the same diagram adopting the subsolar abundances.
Similar to the case of the solar abundances, the calculated
[Fe {\sc vii}]$\lambda$6087/[O {\sc iii}]$\lambda$5007 for BLS1s is 
smaller than that for NLS1s.
In the case of the subsolar abundances, the large fraction of iron is depleted
(see Section 4.2). Therefore, the calculated 
[Fe {\sc vii}]$\lambda$6087/[O {\sc iii}]$\lambda$5007 is much smaller than
that calculated adopting the solar abundances.

\section{CONCLUDING REMARKS}

This paper has presented the comparisons of emission-line ratios 
which represent
the ionization degree and the gas temperature of NLR clouds between
the NLS1s and the BLS1s. The emission-line ratio of 
[O {\sc i}]/[O {\sc iii}]$\lambda$5007, which probes the ionization degree of
NLRs, and that of [O {\sc iii}]$\lambda$4363/[O {\sc iii}]$\lambda$5007,
which probes the gas temperature of NLRs, are indistinguishable
between the two samples. 
This means that there is little difference in the physical properties 
of NLRs between NLS1s and BLS1s.
Using photoionization models, we have confirmed that these results are
consistent with the presence of
differences in SEDs between NLS1s and BLS1s.
In both cases, using the template SEDs of NLS1s and BLS1s,
we have shown that the observed emission  line ratios 
are well reproduced when we adopt
$10^4 {\rm cm}^{-3} \leq n_{\rm H} \leq 10^5 {\rm cm}^{-3}$ and 
$10^{-3.5} \leq U \leq 10^{-3}$
for either solar or subsolar abundances.

This study tells us that we need not consider the effects of 
difference of intrinsic SEDs between NLS1s and BLS1s when we discuss
ionized gas properties using diagnostic diagrams
as used by, e.g., Ferland \& Netzer (1983) and Ho et al. (1993), unless 
the high ionization nuclear emission-line region 
(Binette 1985; Murayama, Taniguchi, \& Iwasawa 1998;
Murayama \& Taniguchi 1998a, 1998b; Nagao et al. 2000b, 2000c) is concerned.

\acknowledgments

We would like to thank the anonymous referee for useful comments and
Gary Ferland for making his code $Cloudy$ available to the public.
This research has made use of the NED (NASA extra galactic database) which is
operated by the Jet Propulsion Laboratory, California Institute of Technology,
under construct with the National Aeronautics and Space Administration.
TM is supported by a Research Fellowship from the Japan Society for the 
Promotion of Science for Young Scientists.
This work was financially supported in part by Grant-in-Aids for the Scientific
Research (Nos. 10044052 and 10304013) of the Japanese Ministry of Education,
Culture, Sports, and Science.

\clearpage

\begin{deluxetable}{lllcl}
\scriptsize
\tablenum{1}
\tablecaption{The Properties of the Objects in Our Sample \label{tbl-1}}
\tablewidth{0pt}
\tablehead{
\colhead{Name} &
\colhead{Another Name} &
\colhead{Redshift} & 
\colhead{$\nu L_{\nu}$(60$\mu$m)} & 
\colhead{References for Emission Line Ratios\tablenotemark{a}} 
}
\startdata
\cutinhead{NLS1s}
NGC 4051 & \nodata                  & 0.0024 & 2.348$\times$10$^{10}$ & M95, V88 \nl
NGC 4748 & MCG -2-33-34, CTS R12.02 & 0.0146 & 1.419$\times$10$^{11}$ & W92, RPD00 \nl
Mrk 42   & \nodata                  & 0.0240 & 1.053$\times$10$^{11}$ & C91, K78, OP85 \nl
Mrk 291  & \nodata                  & 0.0352 & 2.428$\times$10$^{11}$ & O77 \nl
Mrk 335  & PG 0003+199              & 0.0258 & 1.316$\times$10$^{11}$ & C94, O77, P78 \nl
Mrk 359  & UGC 1032                 & 0.0174 & 1.956$\times$10$^{11}$ & C94, DK78, OP85, V88 \nl
Mrk 478  & PG 1440+356              & 0.0791 & 2.161$\times$10$^{12}$ & O77, P78 \nl
Mrk 493  & UGC 10120                & 0.0319 & 4.099$\times$10$^{11}$ & C91, OP85 \nl
Mrk 504  & PG 1659+294              & 0.0359 & \nodata & O77 \nl
Mrk 507  & \nodata                  & 0.0559 & 1.010$\times$10$^{12}$ & K78 \nl
Mrk 739  & NGC 3758                 & 0.0299 & 6.499$\times$10$^{11}$ & C94, SO81 \nl
Mrk 766  & NGC 4253                 & 0.0129 & 3.830$\times$10$^{11}$ & C94, OP85, V88 \nl
Mrk 783  & \nodata                  & 0.0672 & 8.392$\times$10$^{11}$ & OP85 \nl
Mrk 896  & \nodata                  & 0.0264 & 2.068$\times$10$^{11}$ & C94, MW88 \nl
Mrk 957  & 5C 3.100                 & 0.0711 & 6.390$\times$10$^{12}$ & K78 \nl
Mrk 1126 & NGC 7450                 & 0.0106 & \nodata & OP85 \nl
Mrk 1239 & \nodata                  & 0.0199 & 3.038$\times$10$^{11}$ & C91, C94, OP85, RPD00 \nl 
1E 1031+5822    & \nodata           & 0.248 & \nodata & S89 \nl
1E 1205+4657    & \nodata           & 0.102 & \nodata & S89 \nl
1E 12287+123    & \nodata           & 0.116 & \nodata & S89 \nl
2E 1226+1336    & \nodata           & 0.150 & \nodata & S89 \nl
2E 1557+2712    & \nodata           & 0.0646 & \nodata & S89 \nl
I Zw 1          & UGC 545           & 0.0611 & 5.004$\times$10$^{12}$ & C94, O77, P78 \nl
Akn 564         & UGC 12163         & 0.0247 & 2.901$\times$10$^{11}$ & C94, V88 \nl
CTS H34.06      & IRAS F06083-5606  & 0.0318 & 1.396$\times$10$^{11}$ & RPD00 \nl
H 1934-063      & IRAS 19348-0619   & 0.0106 & 1.787$\times$10$^{11}$ & RPD00\nl
HE 1029-1831    & CTS J04.08, IRAS 10295-1831 & 0.0403 & 2.413$\times$10$^{12}$ & RPD00 \nl
IRAS 15091-2107 & \nodata           & 0.0446 & 1.777$\times$10$^{12}$ & W92 \nl
CTS J03.19      & CTS 90            & 0.0532 & \nodata & RPD00 \nl
CTS J13.12      & CTS 103           & 0.0120 & \nodata & RPD00 \nl
KAZ 163         & VII Zw 742        & 0.0630 & \nodata & S89 \nl
KAZ 320         & \nodata           & 0.0345 & \nodata & Z92 \nl
MS 01119-0132   & \nodata           & 0.12 & \nodata & S89 \nl
MS 01442-0055   & \nodata           & 0.08 & \nodata & S89 \nl
MS 12235+2522   & \nodata           & 0.067 & \nodata & S89 \nl
Q 0919+515      & \nodata           & 0.161 & \nodata & S89 \nl
\cutinhead{BLS1s}
NGC 4235 & \nodata                & 0.0080 & 1.158$\times$10$^{10}$ & MW88, M95 \nl
NGC 4593 & \nodata                & 0.0090 & 1.401$\times$10$^{11}$ & MW88 \nl
NGC 5940 & \nodata                & 0.0337 & 4.915$\times$10$^{11}$ & MW88 \nl
Mrk 10   & UGC 4013               & 0.0293 & 4.025$\times$10$^{11}$ & O77 \nl
Mrk 40   & Arp 151                & 0.0211 & \nodata & O77 \nl
Mrk 50   & \nodata                & 0.0234 & \nodata & MW88 \nl
Mrk 69   & \nodata                & 0.0760 & \nodata & O77 \nl
Mrk 79   & UGC 3973               & 0.0222 & 4.250$\times$10$^{11}$ & C83, O77, V88 \nl
Mrk 106  & \nodata                & 0.1235 & \nodata & O77 \nl
Mrk 124  & \nodata                & 0.0563 & 1.286$\times$10$^{12}$ & O77, P78 \nl 
Mrk 141  & \nodata                & 0.0417 & 7.533$\times$10$^{11}$ & O77 \nl
Mrk 142  & PG 1022+519            & 0.0449 & \nodata & O77, P78 \nl
Mrk 236  & \nodata                & 0.0520 & 2.772$\times$10$^{11}$ & O77 \nl
Mrk 279  & UGC 8823               & 0.0294 & 6.275$\times$10$^{11}$ & C83, O77 \nl
Mrk 304  & II Zw 175, PG 2214+139 & 0.0658 & \nodata & O77 \nl
Mrk 358  & \nodata                & 0.0452 & 3.816$\times$10$^{11}$ & O77 \nl
Mrk 374  & \nodata                & 0.0435 & 2.952$\times$10$^{11}$ & O77, P78 \nl
Mrk 382  & \nodata                & 0.0338 & 1.431$\times$10$^{11}$ & O77 \nl
Mrk 486  & PG 1535+547            & 0.0389 & \nodata & O77, P78 \nl
Mrk 541  & \nodata                & 0.0394 & 3.182$\times$10$^{11}$ & O77, P78 \nl
Mrk 590  & NGC 863, UM 412        & 0.0264 & 1.965$\times$10$^{11}$ & O77 \nl
Mrk 618  & \nodata                & 0.0356 & 1.990$\times$10$^{12}$ & O77, P78 \nl
Mrk 704  & \nodata                & 0.0292 & 1.797$\times$10$^{11}$ & C83, C94, UW83 \nl
Mrk 705  & UGC 5025               & 0.0292 & 2.900$\times$10$^{11}$ & C94 \nl
Mrk 876  & PG 1613+658            & 0.1290 & 6.328$\times$10$^{12}$ & S89 \nl
Mrk 975  & UGC 774                & 0.0496 & 1.163$\times$10$^{12}$ & C83, C94, V88 \nl
Mrk 1040 & NGC 931                & 0.0167 & 4.047$\times$10$^{11}$ & C94, V88 \nl
Mrk 1044 & PG 0157+001            & 0.0165 & 6.656$\times$10$^{10}$ & T91 \nl
Mrk 1347 & \nodata                & 0.0497 & 1.337$\times$10$^{12}$ & MW88 \nl
1E 0514-0030 & \nodata            & 0.292 & \nodata & S89 \nl
1H 1836-786  & IRAS F18389-7834   & 0.0743 & 1.246$\times$10$^{12}$ & MW88 \nl
1H 1927-516  & CTS G03.04         & 0.0403 & \nodata & RPD00 \nl
1H 2107-097  & \nodata            & 0.0268 & \nodata & W92, RPD00 \nl
2E 0150-1015 & \nodata            & 0.361 & \nodata & S89 \nl
2E 0237+3953 & \nodata            & 0.528 & \nodata & S89 \nl
2E 1401+0952 & \nodata            & 0.441 & \nodata & S89 \nl
2E 1530+1511 & \nodata            & 0.090 & \nodata & S89 \nl
2E 1556+2725 & PGC 56527          & 0.0904 & \nodata & S89 \nl
2E 1847+3329 & \nodata            & 0.509 & \nodata & S89 \nl
3C 263       & \nodata            & 0.646 & \nodata & P78 \nl
3C 382       & CGCG 173-014       & 0.0579 & \nodata & O77 \nl
III Zw 002   & \nodata            & 0.0898 & \nodata & O77 \nl
VII Zw 118   & \nodata            & 0.0797 & \nodata & C94, KS79 \nl
CTS A08.12   & CTS 109            & 0.0293 & \nodata & RPD00 \nl
AB 125       & \nodata            & 0.281 & \nodata & S89 \nl
Akn 120      & UGC 3271           & 0.0323 & 3.891$\times$10$^{11}$ & P78, V88 \nl
Arp 102B     & \nodata            & 0.0242 & \nodata & SSK83 \nl
CTS B25.02   & Tololo 0343-397    & 0.0432 & 2.622$\times$10$^{11}$ & T91 \nl
B2 1425+26   & Ton 202, PG 1425+267  & 0.366 & \nodata & P78 \nl
B2 1512+37   & 4C 37.43, PG 1512+370 & 0.3707 & \nodata & P78 \nl
CTS C16.16   & \nodata            & 0.0795 & \nodata & RPD00 \nl
ESO 141-G55  & \nodata            & 0.0360 & 4.335$\times$10$^{11}$ & MW88, W92 \nl
ESO 198-G24  & \nodata            & 0.0455 & \nodata & W92 \nl
ESO 323-G77  & \nodata            & 0.0150 & 7.275$\times$10$^{11}$ & W92 \nl
ESO 438-G09  & \nodata            & 0.0245 & 1.088$\times$10$^{12}$ & KF83 \nl
ESO 578-G09  & CTS J14.05         & 0.0349 & \nodata & RPD00 \nl
CTS F10.01   & CTS 114            & 0.0784 & \nodata & RPD00 \nl
Fairall 9    & ESO 113-IG45       & 0.0470 & \nodata & W92 \nl
Fairall 265  & \nodata            & 0.0295 & 3.388$\times$10$^{11}$ & W92 \nl
Fairall 1116 & Tololo 0349-406    & 0.0582 & 3.175$\times$10$^{11}$ & W92 \nl
Fairall 1146 & \nodata            & 0.0316 & \nodata & RPD00 \nl
CTS H34.03   & IRAS F05561-5357   & 0.0967 & 2.769$\times$10$^{12}$ & RPD00 \nl
H 0557-385   & CTS B31.01, IRAS F05563-3820 & 0.0344 & \nodata & W92, RPD00 \nl
IC 4218      & \nodata            & 0.0194 & \nodata & MW88 \nl
IC 4329A     & \nodata            & 0.0161 & 2.987$\times$10$^{11}$ & MW88, W92, WP79 \nl
CTS J10.09   & IRAS F12312-2047   & 0.0230 & 1.164$\times$10$^{11}$ & RPD00 \nl
CTS M02.30   & IRAS F10306-2651   & 0.0688 & 7.266$\times$10$^{11}$ & RPD00 \nl
MC 1104+167  & 4C 16.30           & 0.632 & \nodata & P78 \nl
MS 02255+3121 & \nodata           & 0.058 & \nodata & S89 \nl
MS 07451+5545 & \nodata           & 0.174 & \nodata & S89 \nl
MS 08451+3751 & \nodata           & 0.307 & \nodata & S89 \nl
MS 08495+0805 & \nodata           & 0.062 & \nodata & S89 \nl
MS 10590+7302 & \nodata           & 0.089 & \nodata & S89 \nl
MS 11397+1040 & \nodata           & 0.150 & \nodata & S89 \nl
MS 13396+0519 & \nodata           & 0.266 & \nodata & S89 \nl
MS 13575-0227 & \nodata           & 0.416 & \nodata & S89 \nl
MS 15251+1551 & \nodata           & 0.230 & \nodata & S89 \nl
MS 22152-0347 & \nodata           & 0.242 & \nodata & S89 \nl
PG 1352+183   & \nodata           & 0.152 & \nodata & S89 \nl
PKS 1417-19   & CTS J15.22, CTS 105 & 0.120 & \nodata & RPD00 \nl
Tololo 20     & \nodata           & 0.030 & \nodata & MW88 \nl
Ton 1542      & PG 1229+204       & 0.0640 & \nodata & P78 \nl
Zw 0033+45    & CGCG 535-012      & 0.0476 & \nodata & C94 \nl
\enddata
\tablenotetext{a}{Each abbreviation means as follows; \\
C83   : Cohen (1983) \ \ \ \ \  C91   : Crenshaw et al. (1991) \\
C94   : Cruz-Gonz\'{a}lez et al. (1994) \\
DK78  : Davidson \& Kinman (1978) \ \ \ \ \  K78   : Koski (1978) \\
KF83  : Kollatschny \& Fricke (1983) \\
KS79  : Kunth \& Sargent (1979) \ \ \ \ \  M95   : Murayama (1995) \\
MW88  : Morris \& Ward (1988) \ \ \ \ \  O77   : Osterbrock (1977) \\
OP85  : Osterbrock \& Pogge (1985) \ \ \ \ \  P78   : Phillips (1978) \\
RPD00 : Rodr\'{\i}guez-Ardia, Pastoriza, \& Donzelli (2000b) \\
SSK83 : Stauffer, Schild, \& Keel (1983) \\
S89   : Stephens (1989) \ \ \ \ \  SO81  : Shuder \& Osterbrock (1981) \\
T91   : Terlevich et al. (1991) \\
UW83  : Ulvestad \& Wilson (1983) \\
V88   : Veilleux (1988) \ \ \ \ \  W92   : Winkler (1992)  \\
WP79  : Wilson \& Penston (1979) \ \ \ \ \  Z92   : Zamorano et al. (1992)
}
\end{deluxetable}

\begin{deluxetable}{lccccc}
\tablenum{2}
\tablecaption{Compiled Emission-Line Flux Ratios \label{tbl-2}}
\tablewidth{0pt}
\scriptsize
\tablehead{
\colhead{Name} &
\colhead{[O {\sc i}]/[O {\sc iii}]$\lambda$5007} &
\colhead{[O {\sc ii}]/[O {\sc iii}]$\lambda$5007} & 
\colhead{[O {\sc iii}]$\lambda$4363/[O {\sc iii}]$\lambda$5007} & 
\colhead{[N {\sc ii}]/[O {\sc iii}]$\lambda$5007} & 
\colhead{[S {\sc ii}]/[O {\sc iii}]$\lambda$5007}
}
\startdata
\cutinhead{NLS1s}
NGC 4051 & 0.1300	&	\nodata	&	\nodata	&	1.0190	&	0.3160	\nl
NGC 4748 & 0.0400	&	0.1883	&	0.0492	&	0.3000	&	0.2100	\nl
Mrk 42   & 0.1435	&	0.3696	&	0.1339	&	1.8248	&	0.5680	\nl
Mrk 291  & 0.1031	&	0.4742	&	0.1031	&	1.1031	&	0.4845	\nl
Mrk 335  & 0.0667	&	0.4316	&	0.0699	&   $<$ 1.4783	&	0.0571	\nl
Mrk 359  & 0.0577	&	0.1790	&	0.0735	&	0.2120	&	0.1297	\nl
Mrk 478  & 0.0740	&	0.4280	&	\nodata	&	\nodata	&	0.2040	\nl
Mrk 493  & \nodata	&	0.5600	&	0.4568	&	2.0400	&	0.8240	\nl
Mrk 504  & \nodata	&	\nodata	&	0.0308	&	\nodata	&	0.1192	\nl
Mrk 507  & 0.2099	&	0.5679	&	0.0593	&	4.3951	&	1.1481	\nl
Mrk 739  & 0.3700	&	0.8894	&   $<$ 0.0740	&	4.2800	&	2.1500	\nl
Mrk 766  & 0.0241	&	0.0763	&	0.0294	&	0.2166	&	0.0695	\nl
Mrk 783  & 0.0584	&	0.2957	&	0.0895	&	0.1479	&	0.1479	\nl
Mrk 896  & \nodata	&	0.5276	&	\nodata	&	\nodata	&	0.1626	\nl
Mrk 957  & 0.3016	&	0.8730	&	0.0952	&	3.4444	&	1.1111	\nl
Mrk 1126 & 0.0791	&	0.1279	&	0.1163	&	5.8140	&	0.3233	\nl
Mrk 1239 & 0.0200	&	0.0852	&	0.0892	&	0.3011	&	0.1128	\nl
1E 1031+5822    & 1.1280	&	0.7082	&	0.0683	&	\nodata	&	\nodata	\nl
1E 1205+4657    & 0.0555	&	\nodata	&	0.3646	&	\nodata	&	0.7376	\nl
1E 12287+123    & \nodata	&	\nodata	&	0.5395	&	\nodata	&	0.6974	\nl
2E 1226+1336    & \nodata	&	0.3493	&	0.0312	&	\nodata	&	0.4521	\nl
2E 1557+2712    & \nodata	&	\nodata	&	\nodata	&	\nodata	&	0.6269	\nl
I Zw 1          & 0.0636	&	0.4756	&	0.0761	&	\nodata	&	0.0295	\nl
Akn 564         & \nodata	&	0.4068	&	\nodata	&	\nodata	&	\nodata	\nl
CTS H34.06      & \nodata	&	0.3421	&	0.1023	&	\nodata	&	\nodata	\nl
H 1934-063      & \nodata	&	0.1257	&	0.0437	&	\nodata	&	\nodata	\nl
HE 1029-1831    & \nodata	&	0.1961	&	0.0980	&	\nodata	&	\nodata	\nl
IRAS 15091-2107 & \nodata	&	0.1400	&	0.0500	&	0.3200	&	0.3600	\nl
CTS J03.19      & \nodata	&	0.1626	&	\nodata	&	\nodata	&	\nodata	\nl
CTS J13.12      & \nodata	&	\nodata	&	0.3101	&	\nodata	&	\nodata	\nl
KAZ 163         & \nodata	&	\nodata	&	\nodata	&	\nodata	&	0.1418	\nl
KAZ 320         & 0.0629	&	0.3452	&	0.0595	&	0.2151	&	0.2738	\nl
MS 01119-0132   & \nodata	&	\nodata	&	0.1188	&	\nodata	&	\nodata	\nl
MS 01442-0055   & \nodata	&	\nodata	&	0.0207	&	\nodata	&	0.2434	\nl
MS 12235+2522   & \nodata	&	\nodata	&	\nodata	&	\nodata	&	0.2120	\nl
Q 0919+515      & \nodata	&	0.2191	&	\nodata	&	\nodata	&	\nodata	\nl
\cutinhead{BLS1s}
NGC 4235 & 0.1967	&	0.6389	&	0.2837	&	0.6610	&	0.6252	\nl
NGC 4593 & 0.0470	&	0.1696	&	\nodata	&	\nodata	&	0.1357	\nl
NGC 5940 & 0.0772	&	\nodata	&	\nodata	&	\nodata	&	0.1680	\nl
Mrk 10   & 0.0172	&	0.0871	&	0.0247	&	0.1183	&	0.0763	\nl
Mrk 40   & 0.1111	&	0.5000	&	0.1667	&	0.2639	&	0.1444	\nl
Mrk 50   & 0.1218	&	0.3261	&	\nodata	&	\nodata	&	0.1851	\nl
Mrk 69   & \nodata	&	0.4186	&	0.0837	&	1.0698	&	0.3512	\nl
Mrk 79   & 0.0424	&	0.1763	&	0.0457	&	0.2591	&	0.1741	\nl
Mrk 106  & 0.0414	&	0.2000	&	0.0310	&	0.2241	&	\nodata	\nl
Mrk 124  & 0.0628	&	0.3577	&	0.0730	&	\nodata	&	0.2394	\nl
Mrk 141  & \nodata	&	0.1632	&	0.0816	&	0.9474	&	0.3763	\nl
Mrk 142  & $<$ 0.0857	&	0.3760	&	0.0680	&	1.3200	&	0.5669	\nl
Mrk 236  & 0.0717	&	0.1848	&	0.0804	&	0.1587	&	0.2565	\nl
Mrk 279  & 0.1312	&	0.4635	&	0.0670	&	0.6483	&	0.3204	\nl
Mrk 304  & $<$ 0.0455	&	0.1212	&	0.1061	&	0.3788	&	0.1818	\nl
Mrk 358  & $<$ 0.0977	&	0.2326	&   $<$ 0.0605	&	0.3256	&	0.3953	\nl
Mrk 374  & 0.0686	&	0.0914	&	0.0738	&	0.0628	&	0.1188	\nl
Mrk 382  & \nodata	&	0.0805	&	0.0322	&	0.3333	&	0.7805	\nl
Mrk 486  & 0.0917	&   $<$ 0.1500	&	\nodata	&	\nodata	&	0.0773	\nl
Mrk 541  & $<$ 0.2050	&	0.1739	&	0.2217	&	1.3755	&	0.5600	\nl
Mrk 590  & 0.2000	&	0.2182	&	0.2000	&	0.5455	&	0.2345	\nl
Mrk 618  & 0.0549	&	0.1356	&	0.0410	&	0.8667	&	0.2320	\nl
Mrk 704  & 0.0287	&	0.2016	&	0.0726	&	0.1710	&	0.1507	\nl
Mrk 705  & \nodata	&	0.1532	&	\nodata	&	\nodata	&	\nodata	\nl
Mrk 876  & 0.1659	&	\nodata	&	0.0272	&	\nodata	&	0.7093	\nl
Mrk 975  & 0.0248	&	0.1998	&	0.0303	&	0.4421	&	0.0634	\nl
Mrk 1040 & \nodata	&	0.0850	&	\nodata	&	\nodata	&	\nodata	\nl
Mrk 1044 & \nodata	&	0.1238	&	\nodata	&	\nodata	&	\nodata	\nl
Mrk 1347 & \nodata	&	0.6032	&	\nodata	&	\nodata	&	0.2063	\nl
1E 0514-0030 & \nodata	&	\nodata	&	0.4949	&	\nodata	&	\nodata	\nl
1H 1836-786  & 0.0349	&	0.1623	&	\nodata	&	\nodata	&	0.1623	\nl
1H 1927-516  & \nodata	&	\nodata	&	0.0792	&	\nodata	&	\nodata	\nl
1H 2107-097  & 0.1300	&	0.1300	&	0.1696	&	\nodata	&	\nodata	\nl
2E 0150-1015 & \nodata	&	0.1291	&	0.0341	&	\nodata	&	\nodata	\nl
2E 0237+3953 & \nodata	&	0.3364	&	0.0757	&	\nodata	&	\nodata	\nl
2E 1401+0952 & \nodata	&	\nodata	&	0.0565	&	\nodata	&	\nodata	\nl
2E 1530+1511 & \nodata	&	\nodata	&	\nodata	&	\nodata	&	0.2091	\nl
2E 1556+2725 & 0.0430	&	\nodata	&	0.0884	&	\nodata	&	0.2033	\nl
2E 1847+3329 & \nodata	&	0.1800	&	0.0649	&	\nodata	&	\nodata	\nl
3C 263       & \nodata	&	\nodata	&	0.0515	&	\nodata	&	\nodata	\nl
3C 382       & 0.0700	&	0.1300	&	0.1600	&	0.0400	&	0.1600	\nl
III Zw 002   & 0.0576	&	0.1394	&	0.0273	&	0.1333	&	0.1576	\nl
VII Zw 118   & \nodata	&	0.4444	&	\nodata	&	\nodata	&	\nodata	\nl
CTS A08.12   & \nodata	&	0.1280	&	0.0759	&	\nodata	&	\nodata	\nl
AB 125       & \nodata	&	0.2922	&	\nodata	&	\nodata	&	\nodata	\nl
Akn 120      & 0.0917	&	0.2658	&	\nodata	&	\nodata	&	0.3108	\nl
Arp 102B     & 0.8030	&	1.1970	&	0.0571	&	1.9343	&	1.2475	\nl
CTS B25.02   & \nodata	&	0.3207	&	\nodata	&	\nodata	&	\nodata	\nl
B2 1425+26   & \nodata	&	\nodata	&	0.0417	&	\nodata	&	\nodata	\nl
B2 1512+37   & \nodata	&	\nodata	&	0.0393	&	\nodata	&	\nodata	\nl
CTS C16.16   & \nodata	&	0.1793	&	0.0588	&	\nodata	&	\nodata	\nl
ESO 141-G55  & \nodata	&	0.0924	&	\nodata	&	\nodata	&	0.2800	\nl
ESO 198-G24  & \nodata	&	0.2700	&	\nodata	&	\nodata	&	0.1300	\nl
ESO 323-G77  & \nodata	&	0.1300	&	\nodata	&	\nodata	&	0.2400	\nl
ESO 438-G09  & 0.0811	&	0.4324	&	0.0541	&	1.0000	&	0.4595	\nl
ESO 578-G09  & \nodata	&	0.1719	&	0.0898	&	\nodata	&	\nodata	\nl
CTS F10.01   & \nodata	&	0.1048	&	0.0391	&	\nodata	&	\nodata	\nl
Fairall 9    & \nodata	&	0.0800	&	0.1200	&	0.2000	&	\nodata	\nl
Fairall 265  & \nodata	&	0.7600	&	\nodata	&	1.3500	&	\nodata	\nl
Fairall 1116 & \nodata	&	0.2000	&	0.0900	&	\nodata	&	0.5400	\nl
Fairall 1146 & \nodata	&	0.1570	&	0.0436	&	\nodata	&	\nodata	\nl
CTS H34.03   & \nodata	&	\nodata	&	0.1287	&	\nodata	&	\nodata	\nl
H 0557-385   & \nodata	&	0.0816	&	\nodata	&	\nodata	&	0.1200	\nl
IC 4218      & \nodata	&	0.2318	&	\nodata	&	\nodata	&	0.1706	\nl
IC 4329A     & 0.0964	&	0.0429	&	\nodata	&	0.3091	&	0.3156	\nl
CTS J10.09   & \nodata	&	0.1034	&	0.1565	&	\nodata	&	\nodata	\nl
CTS M02.30   & \nodata	&	0.1597	&	0.1875	&	\nodata	&	\nodata	\nl
MC 1104+167  & \nodata	&	\nodata	&	0.0941	&	\nodata	&	\nodata	\nl
MS 02255+3121 & \nodata	&	\nodata	&	0.1270	&	\nodata	&	\nodata	\nl
MS 07451+5545 & \nodata	&	0.2066	&	0.0337	&	\nodata	&	\nodata	\nl
MS 08451+3751 & \nodata	&	\nodata	&	0.1198	&	\nodata	&	\nodata	\nl
MS 08495+0805 & 0.0313	&	\nodata	&	0.0395	&	\nodata	&	0.1261	\nl
MS 10590+7302 & \nodata	&	\nodata	&	\nodata	&	\nodata	&	0.1080	\nl
MS 11397+1040 & \nodata	&	\nodata	&	0.0215	&	\nodata	&	0.1304	\nl
MS 13396+0519 & \nodata	&	0.5666	&	\nodata	&	\nodata	&	\nodata	\nl
MS 13575-0227 & \nodata	&	0.1029	&	\nodata	&	\nodata	&	\nodata	\nl
MS 15251+1551 & \nodata	&	0.0675	&	0.0405	&	\nodata	&	\nodata	\nl
MS 22152-0347 & \nodata	&	0.1071	&	0.0329	&	\nodata	&	\nodata	\nl
PG 1352+183   & \nodata	&	\nodata	&	\nodata	&	\nodata	&	0.2090	\nl
PKS 1417-19   & \nodata	&	\nodata	&	0.1039	&	\nodata	&	\nodata	\nl
Tololo 20     & 0.0425	&	0.0800	&	0.0924	&	\nodata	&	0.0337	\nl
Ton 1542      & \nodata	&	0.1944	&	0.1167	&	\nodata	&	0.1333	\nl
Zw 0033+45    & 0.2727	&	\nodata	&	\nodata	&	\nodata	&	\nodata	\\
\enddata
\end{deluxetable}

\clearpage

\end{document}